\documentstyle[prb,aps,twocolumn,floats]{revtex}
\input epsf

\input epsf.sty

\def\prb{Phys. Rev. B}
\def\prl{Phys. Rev. Lett.}

\def\be{\begin{equation}}
\def\ee{\end{equation}}
\def\ba{\begin{eqnarray}}
\def\ea{\end{eqnarray}}

\def\YBCO{YBa$_2$Cu$_3$O$_{7-\delta}$}

\def\C60{A$_x$C$_{60}$}

\def\ie{ {\it i.e.} }

\begin{document}

\twocolumn[\hsize\textwidth\columnwidth\hsize\csname@twocolumnfalse\endcsname

\title
{Doped Antiferromagnets in High Dimension}

\author{E.~W.~Carlson, S.~A.~Kivelson, and Z.~Nussinov}
\address
{Dept. of Physics,
U.C.L.A.,
Los Angeles, CA  90024}
\author{V.~J.~Emery}
\address{
Dept. of Physics,
Brookhaven National Laboratory,
Upton, NY  11973-5000}
\date{\today}
\maketitle 

\begin{abstract}

The ground-state properties of the $t-J$ model on a $d$-dimensional 
hypercubic lattice are examined in the limit of large $d$.  It is found that 
the undoped system is an ordered antiferromagnet, and that the doped 
system phase separates into a hole-free antiferromagnetic phase 
and a hole-rich 
phase.  The latter is electron free if $J > 4t$ and is weakly metallic 
(and typically superconducting) if $J < 4t$.   
The resulting phase diagram is qualitatively similar to the one previously
derived for $d=2$ by a combination of  
analytic and numerical methods. Domain wall (or stripe) phases form
in the presence of weak Coulomb interactions, 
with periodicity determined by the hole concentration and
the relative strength of the exchange and Coulomb interactions.  
These phases
reflect the properties of the hole-rich phase in the absence  of Coulomb
interactions, and,
depending on the value of $J/t$,
may be either insulating or
metallic ({\it i.e.} an ``electron smectic'').

\

\

\end{abstract}

]


In this paper, the zero-temperature properties of the $t-J$ model of a doped 
antiferromagnet on a $d$-dimensional hypercubic lattice are evaluated using
a systematic expansion in powers of $1/d$.  
For each property of interest the leading behavior in the 
large $d$ 
limit is computed, and in some cases, just to prove how tough we are,
corrections up to order $1/d^5$   are obtained. These results are obtained by breaking the full
Hamiltonian into an unperturbed piece, $H_0$, and
a perturbation, $H_1$, and then reorganizing conventional perturbation 
theory in powers of $H_1$ into a $1/d$ expansion.  
Of course the partition of the Hamiltonian may
be chosen for calculational convenience, since it does not affect the results.  
The convergence of this expansion will not be addressed, although we 
believe it to be only asymptotic. 

Our procedure differs from the extensive recent work on the related problem 
of the Hubbard and Falicov-Kimball models in large dimension\cite{kotliar} in
the way the large dimension limit is taken. First of all, we do not assume 
that the ratio $J/t$ of the exchange integral $J$ and the hopping amplitude 
$t$ is parametrically small as $d\rightarrow \infty$. The previous studies 
assumed that $t$ is proportional to $1/\sqrt{d}$ so that, when $J$ is expressed in 
terms of the onsite interaction $U$, it follows that
$J/t = 4t/U \sim 1/\sqrt{d}$. (The phase diagram will
be studied for parametrically 
small values of $J/t$ in Sec. VIII, but our 
results are less complete in this case, because of the difficulty of 
controlling perturbation theory in this limit.)
Secondly, the hypercubic lattice is bipartite, {\it i.e.} it can be broken 
into two sublattices, which we label ``black'' and ``red'', such that the 
Hamiltonian has interactions only between sites on different sublattices. This  
favors the classical N\'eel state, which has a uniaxial magnetization with
opposite sign on the two sublattices. By contrast, earlier studies, which  
were primarily concerned with the Mott transition and possible non-Fermi 
liquid states of the Hubbard model, assumed a non-bipartite lattice
which frustrates the N\'eel state.  For both reasons, this previous work
does not shed much light on the behavior of
doped antiferromagnets. 
A notable exception is the work of van Dongen\cite{dongen} on the small $U$ limit 
of the Hubbard model on a hypercubic lattice which found, as we do, that the 
weakly-doped
antiferromagnetic phase is unstable to phase separation, even if the 
parameters are scaled as $J^{eff}/t = 4t/U \sim 1/\sqrt{d}$.
 
Throughout this paper, units are chosen such that the lattice constant,
$\hbar$, and Boltzmann's constant are all equal to one.
 
\section{  Summary of Results}

\subsection{Results in Large Dimension}

Our principal result is the global zero-temperature phase 
diagram as a function of $J/t$ and hole concentration $x$, in the limit of 
large $d$, as shown in Fig. 1.  It is immediately clear that in most of the 
phase diagram, the undoped (ordered) antiferromagnetic phase coexists with a 
hole-rich phase.  For $J/t > 4$, the hole-rich phase is electron free; 
otherwise it contains an exponentially small but non-vanishing
concentration of electrons.  In the intermediate coupling regime,
$2 < J/t < 4$, the residual attraction ( $J$ ) between electrons is 
great enough to overcome the hard-core repulsion, and leads to a 
BCS instability
of the dilute metal, producing an s-wave superconducting state at 
exponentially low energy scales.  At smaller values of $J/t$, the net
interaction between electrons is repulsive.  This implies that the
system either remains metallic down to zero temperature or
exhibits higher-angular-momentum pairing \cite{hm2} via 
the Kohn-Luttinger mechanism.\cite{KL}

\begin{figure}[htb]
\centerline{\epsfxsize=3.3in \epsffile{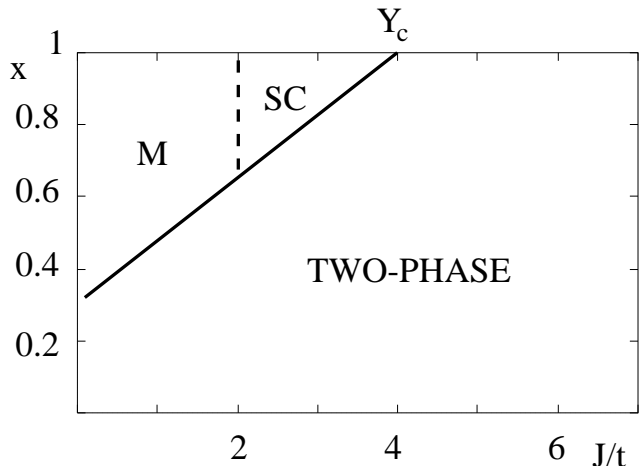}}
\caption{Phase diagram of the $t-J$ model in the limit $d\rightarrow \infty$:
Here $x$ is the hole concentration ($1-x$ is the electron concentration).
The phase boundary is given by Eq. (\protect \ref{eq:nmax}), artificially 
setting $d=2$.  
``Two-phase'' labels the two-phase region, where a uniform density phase is 
thermodynamically unstable, ``SC'' labels a region of $s-$ wave 
superconductivity, and
``M'' labels a region of metallic behavior with 
repulsive interactions, which presumably has an ultra-low temperature
superconducting instability due to the Kohn-Luttinger effect.\protect \cite{KL}}
\label{fig1}
\end{figure}

A peculiarity of the phase diagram in Fig. 1 is that the boundary of the
two-phase region intersects the $J/t=0$ axis at a non-zero value of $x$.
This is not likely to be correct in any finite dimension.  For small $x$ and large
but finite dimension,  we
expect  that in the limit $J/t\rightarrow 0$, the ground state is a ferromagnetic
Fermi liquid, and hence the model does not phase separate.  In Section 
 VIII, we
discuss the behavior of the model for $J/t$ parametrically small,
$J/t \sim 1/\sqrt{d}$.  Here the
$1/d$ expansion is slightly more difficult to control, so our 
results, summarized in Fig. 2, are incomplete.  
The resulting conjectural phase diagram for large but finite $d$ embodies
all the insights gained from studying the $d \rightarrow \infty$ limit,
but corrects the unphysical features of the phase diagram in Fig. 1.

\begin{figure}[htb]
\centerline{\epsfxsize=3.3in \epsffile{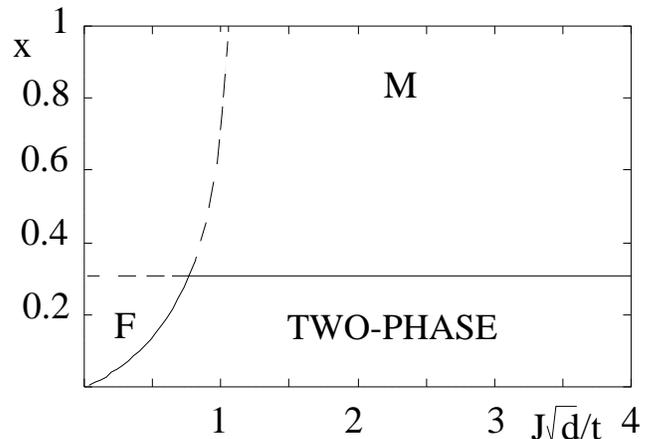}}
\caption{Conjectured Phase diagram of the $t-J$ model 
for large but finite $d$:  This figure should be viewed as a blowup of
the small $J/t$ portion of Figure 1.  The horizontal line
represents the small $J/t$ extension of the phase boundary in Fig. 1;
in fact, in large $d$, this line would be exponentially close to
the top of the figure, but we have drawn it, as in Fig. 1,
at a position obtained by setting $d=2$ in the large $d$ expression.
The boundary of the fully polarized ferromagnetic metallic
phase (labelled ``F'') 
is drawn in accord with the large $d$ expression  in Eq. 
(\protect \ref{eq:nmaxferro}. There might be other,
lower energy phases, ({\it e.g.} high-density stripe phases)
that could occur below these two phase boundaries,
in the region marked ``two-phase'',
especially close to the point of intersection.}
\label{fig2}
\end{figure}

We have also studied the behavior of one or two doped holes and the character 
of charged domain walls in the antiferromagnet. It will be seen that the 
latter are stablilized by a long-range Coulomb interaction. These studies
bring out an important characteristic of our large-$d$ expansion. Whenever
a hole moves in the antiferromagnetic background, it may break a number of 
bonds of order $d$ at each hop. Consequently, {\it for such processes}, 
the physics is exchange dominated for large-$d$ and it amounts to an expansion 
in powers of
$t/J$. This is true of the
motion of one or two holes and of domain wall fluctuations in which holes hop
into the environment. However the questions of phase separation, domain wall 
phase equilibrium, and superconductivity at low electron concentration are 
not subject to this limitation. 

The ensuing discussion will be organized by 
order of increasing hole concentration.

To leading order in $1/d$ 
the states of minimum energy of a single hole lie precisely on the magnetic 
Brillouin zone which also is the Fermi surface of the noninteracting 
system with a half-filled band.  The massive degeneracy of these low energy
hole states is lifted by terms of ${\cal O}(1/d^4)$, and it is found that
the absolute minimum occurs at $\vec k=(\pi/2)<1,1,1, ...>$ together with 
points related by the point-group symmetry.
Moreover, as deduced previously  by Trugman\cite{trugman} in studies of
two holes in a two-dimensional antiferromagnet, we find that propagation
of pairs of holes is no less frustrated 
than is the propagation of a single hole, because of a subtle effect
of Fermi statistics.  There is, however,
an effective attraction $\sim 1/d$ between two holes due to the fact that 
two nearest-neighbor holes break one less antiferromagnetic
bond than two far-separated holes;  this attraction always leads to a 
two-hole bound state.

An interesting metastable state is 
a charged magnetic domain wall ({\it i.e.} a $d-1$ 
dimensional hypersurface with finite hole concentration and suppressed
magnetic order). We have found that the most stable
domain wall has an electron density which is,
to leading order in $1/d$, equal to that of the hole-rich phase which
can exist in equilibrium with the antiferromagnet.  Thus domain walls
can be viewed as a form of local phase separation.  Also
the domain wall configuration with the lowest surface tension
({\it i.e.} energy per unit hyperarea of wall) is the ``vertical'' 
site-centered $\pi$ (antiphase) discommensuration in the antiferromagnetic 
order; {\it i.e.} it is parallel to a single nearest-neighbor vector and odd 
under reflection through a site-centered vertical hyperplane.

We have considered the effect of weak, long-range Coulomb interactions as a
perturbation.  While this study is not exhaustive, we conclude that, for a 
substantial range of parameters, the ground state consists of a 
periodically-ordered array of optimal domain walls of the sort described 
above, especially when $x$ is small but not too small. In this range of $x$, 
the ground state is insulating for $J/t>4$, and metallic for $J/t<4$.  The 
latter phase is an ``electron-smectic'' \cite{smectic} which 
exhibits crystalline order in one direction and liquid-like behavior in the
transverse ($d-1$) directions. The liquid features are
associated primarily with the motion of electrons along the domain wall,
and they may be metallic or 
condensed into a superconducting state.

We have argued previously that the
competition between a local tendency to phase separation in a doped
antiferromagnet and the long-range Coulomb repulsion between holes produces 
a large variety of intermediate scale structures, including arrays of
domain walls, which are significant features of doped antiferromagnets that
we have called ``frustrated phase separation''  \cite{frustrated,losalamos}. 
However, these  phenomena have not previously been derived from a 
microscopic {\it magnetic} model.\cite{grilli} 
It is particularly striking that,
in the appropriate range of parameters, charge and spin density wave order 
coexist with metallic, and even superconducting behavior.
 
\subsection{How Large Are $d=2$ and $d=3$?}

Large $d$ is, of course, only of academic interest;  we are interested in
the physical dimensions, $d=1$, 2, and 3.  The properties of the one-dimensional
electron gas (1DEG) are well understood\cite{1deg} by now, and exhibit 
behavior that is quite dimension specific.  Moreover, for 
most of the conceivable ordered states, the lower critical dimension for 
long range order at zero temperature is one, so the 1DEG is not likely to be 
well understood in terms of adiabatic continuity from large dimension.  
However, long range order at zero temperature is quite robust
in both two and three dimensions, so there is every reason to expect that
a $1/d$ expansion will capture the essential physics of 
many of the zero-temperature thermodynamic states. 

To test this conjecture, we would like to make both
qualitative and quantitative comparisons between the results of the large
$d$ theory and any available exact, or well-controlled numerical or analytic 
results in two and three dimensions.  Table 1 gives a quantitative
comparison between the $1/d$ expansion and well-established numerical results 
for the undoped system, {\it i.e.} for the spin-1/2 Heisenberg 
antiferromagnet.  It can be seen that the ground-state energy 
can be obtained from the low-order expansion in powers of $1/d$ to $0.6\%$ 
accuracy or better. By carrying the series to higher order, and possibly 
doing a Pad\'e analysis of the series much improved accuracy for all physical
quantities could be expected. In Sec. XI comparison will be made between 
numerical results and the results of perturbation
theory about the Ising limit (Table 2).

\begin{center}
\begin{tabular}{|c|c|c|c|c|c|}
\hline
{} & {$E_{AF}(2)$} & 
{$m(2)$} 
 & {$E_{AF}(3)$} & {$m(3)$} & {$E_{2-leg}(2)$}  \\
\hline
\hline
{$d^0$} & {-0.5} &
{0.5} & {-0.75} & {0.5} & {-0.25} \\
\hline
{$d^{-1}$} & {-0.625}
& {0.4375} & {-0.875} & {0.4583} &{-0.5}\\
\hline
{$d^{-2}$} & {-0.6563} & {0.4063} & {-0.8958} & {0.4444} & {-0.5625}
 \\
\hline
{$d^{-3}$} & {-0.6631} & {0.3948} & {-0.8989} & {0.4410} & {-0.5664}
\\
\hline
{$d^{-4}$} & {-0.6647}
 & {0.3903} & {-0.8993} & {0.4402} & {-0.5713}\\
\hline
\hline
{Exact} & {-0.669} & 
{0.307} & { - } & { - } & {-0.5780\cite{2leg}}\\
\hline
{Upper} & {-0.5} & { - } & {-0.75 } & { - } & {-0.375} \\
\hline
{Lower} & {-0.75} & { - } & {-1.0} & { - } & {-0.625} \\
\hline
\hline
\end{tabular}
\end{center}

Table 1:  Comparison of the results of exact numerical studies\cite{2d}
(the row labeled ``exact'') on the two-dimensional 
spin-1/2 Heisenberg antiferromagnet 
with the perturbative results in powers of $1/d$ derived in the present paper.
(We have been unable to find corresponding ``exact'' three dimensional
results.)
The dimension is indicated by the arguments of the computed quantities. 
The rows labeled ``Upper'' and ``Lower'' give the rigorous upper and lower 
bounds on the energies obtained in the text.  The approximate results
are obtained by setting $y\equiv J_{\perp}/J_z =1$, $V=0$, and $d=2$ or $3$ in the 
series expansion, evaluated to the stated order.  All energies are measured 
in units of $J/d$, and the magnetization $m$ is quoted in units in which 
$g\mu_B=1$, where $\mu_B$ is the Bohr magneton. 
\vspace*{1cm}

\begin{figure}[htb]
\centerline{\epsfxsize=3in \epsffile{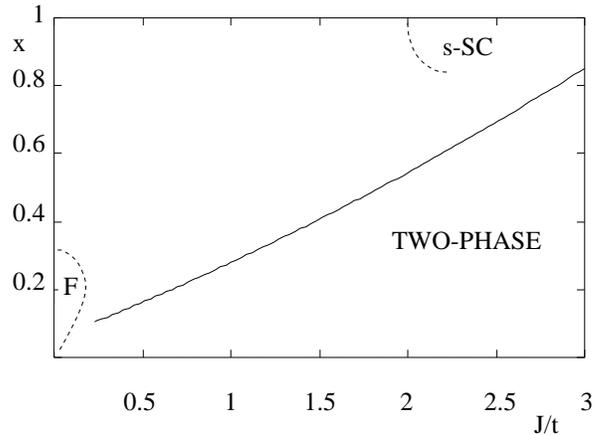}}
\caption{Zero Temperature Phase Diagram of the two dimensional 
$t-J$ model, deduced from numerical studies of finite size systems with up to 
60 electrons, as well as from various analytic results.  
This figure is abstracted from
Hellberg and Manousakis.\protect \cite{hm}}
\label{fig3}
\end{figure}

Qualitative comparisons can be made with the phase diagram of the
two dimensional $t-J$ model which has been deduced from combined analytic
and numerical\cite{ekl,hm} studies.  Figure 3, abstracted from
the work of Hellberg and Manousakis\cite{hm} shows the phase diagram deduced
from numerical studies of systems with up to 60 electrons.  As in large $d$,
there is no thermodynamically stable zero-temperature phase with dilute
holes for any $J/t$.  Indeed, aside from the behavior of the boundary of the
two-phase region at very small $J/t$, the phase diagrams in Figs. 1 and 3 are
similar.  As suggested above,
when the pathologies of the formal $d\rightarrow \infty$ limit
are removed by taking into account the
new processes that become important at parametetrically small values
of $J/t \sim d^{-1/2}$, one obtains for large but finite
$d$ the phase diagram shown in  Figure 2, which is topologically
equivalent to Figure 3.  (Of course, in $d=2$, parametrically
small values of $J/t$ are not all that small, so there is no reason
to expect quantitative agreement with the large $d$ results.  
The critical
value of $J/t=Y_c$ at which the phase-coexistence line deviates from
$x=1$ is\cite{hm} $Y_c=3.4367$ in $d=2$, and is rather well approximated
by the value $Y_c \rightarrow 4$
as $d\rightarrow \infty$.  However, the slope of the phase coexistence line in
$d=2$ is much steeper than would be deduced from the large $d$ theory.)
Similar detailed information
on the three dimensional $t-J$ model is not available at this time, although
arguments presented hitherto \cite{frustrated}, \cite{ekl} suggest that the phase diagram
is qualitatively similar to that in $d=2$, consistent with the
expectations from the $1/d$ expansion.

Our calculation of the spectrum of one hole in an antiferromagnet 
may be compared to the numerical calculations of 
Dagotto {\it et al.}
\cite{onehole} on 
$d=2$ systems with
16$\times$16 sites and 
$J/t=0.4$.  They found that
the one-hole spectrum is well represented by the two-dimensional version
of the expression in Eq. (\ref{eq:hole}), confirming the qualitative
accuracy of the large $d$ expression.  However
the values of the parameters obtained to leading order in $1/d$
are quantitatively quite far from the
exact results, and this discrepancy is made worse by the inclusion of higher 
order terms. This is not unexpected in view of the fact that large $d$
drives the motion of a single hole into the exchange-dominated limit.
In particular, it is clear from Eq. (\ref{eq:wpos}) that the large-$d$
expansion gives a negative value for the bandwidth $W$ in $d=2$, unless
$J > 0.93 t$.
Thus it is essential to compare the large-$d$ expansion 
to numerical results at large $J/t$. 
Specifically, from Eq. (\ref{eq:bandwidth}) with $y=1$, the bandwidth for
$d=2$ is given by $dW/t = 2.125t/J - 1.83(t/J)^3$. It would be interesting
to compare this result with numerical calculations for large $J/t$ extrapolated 
to the thermodynamic limit. Martinez and Horsch\cite{largeJ} have found that
an approximate treatment of the motion of a single hole gives $dW/t = 2t/J$
for large $J/t$, which agrees very well with our large-$d$ result.
Via a variational calculation, Bonisegni and Manousakis\cite{bonisegni}
find $dW/t = 0.59 \pm 0.15$ in the thermodynamic limit for $d=2$ and $J/t = 5$,
while Eq. (\ref{eq:bandwidth}) gives 0.41.

Finally,  we can extrapolate to two dimensions the character of the ordered 
arrays of charged domain walls 
at low doping concentration and weak Coulomb interaction.
Domain walls in two dimensions are one-dimensional
(lines) and such ordered arrays are known as ``stripe phases''.
Directly extrapolating the optimal large dimensional domain-wall structures to
$d=2$, we would expect the stripes to be site-centered, 
vertical, antiphase domain
walls in the antiferromagnetic order, and to be metallic (and possibly
superconducting) for $J/t < Y_c$ and insulating for $J/t> Y_c$.
In particular, if we extrapolate the leading order
expression for the electron density in the hole-rich phase,
Eq. (\ref{eq:nmaxmax}), to $d=2$, and then evaluate it for $t \gg J$,
we find that such stripes should have approximately 0.31 doped holes per site
along the stripe, and are thus metallic.
Transverse to the stripe direction, 
such a phase is a generalized charge and spin 
density wave
state, in which the period of the charge density wave is half that
of the spin density wave.\cite{landau}  However, because of the
electronic motion along the stripe, this phase is actually an electron smectic.
\cite{smectic}
Unfortunately, there are no detailed microscopic two
dimensional calculations to compare with these results, so we
compare them with experiments on doped 
antiferromagnets.\cite{whitescal}

\subsection{Rigorous Results}

In addition to our perturbative results in powers of $1/d$,
we have obtained 
rigorous upper and lower bounds on the ground-state energy of the undoped system.
These bounds, which are also quoted in Table 1,
are shown to converge in the limit
$d\rightarrow \infty$.

\subsection{Relation to Experimental Results on Doped
Antiferromagnets in Quasi-Two and Three
Dimensions}

By now there are many examples of antiferromagnetic insulators that can
be chemically doped.
One prominent feature of these materials is the occurrence of high temperature
superconductivity, a phenomenon for which the present results provide little
{\it direct} insight \cite{spingap}.  However various spin and charge ordered 
states,  as well as ``nearly ordered'' fluctuating versions of such structures,
have been observed in these systems \cite{stripes,cupr} by direct structural 
probes, especially neutron scattering.
Two concrete,
and well studied examples of this are the quasi two-dimensional
perovskites La$_{2-x}$Sr$_x$NiO$_4$ and La$_{1.6-x}$Nd$_x$Sr$_x$CuO$_4$,
in both of which 
the doped hole concentration is
equal to the Sr concentration $x$. 
The undoped parent compounds (with $x=0$) are
antiferromagnetic insulators with spin $S=1$ for the nickelates and
$S=1/2$ for the cuprates.
In both cases, upon doping,
the system forms\cite{nick,cupr}
 a ``stripe'' phase, in which the doped holes are
concentrated in antiphase domain walls in the antiferromagnetic order.
At present it is not known whether the domain walls are site or bond
centered in general.  (At higher doping concentration in the
nickelates, there is strong evidence that both types of domain wall coexist 
due to interactions between the walls \cite{sitecentered}.)  However,
there is a crucial difference between the domain walls in the
two materials:  In the nickelates, there is one doped hole per site along
the domain walls, and the doped system is, correspondingly, insulating.
In the cuprate, the hole concentration along the domain wall is roughly
one doped hole per two sites along the domain wall, and the
system is correspondingly metallic, and even superconducting, despite
the presence of almost static charge and spin density wave order.  (This latter
behavior
is very suggestive evidence of an electron-smectic phase.\cite{smectic})
In addition, 
the domain walls are diagonal in the nickelates\cite{nick}
and vertical in the
cuprates.\cite{cupr,ybco}


We feel that the occurrence of charged stripes
in lightly-doped antiferromagnets,
the fact that these stripes are antiphase domain walls in the
antiferromagnetic order, and that they can be metallic or insulating,
depending on the ratio of $J/t$, are physically robust features of
the large $d$ theory which we expect to apply {\it mutatis mutandis} in $d=2$.
However, the preference for vertical versus diagonal stripes,
and site-centered versus bond-centered stripes is likely to depend
on microscopic details, even in large dimensions.
Of more profound importance is the fact that, while in large
dimensions the charged domain walls always crystallize
at low temperature into an ordered density wave, in low dimensions, especially
in two dimensions, there is the very real possibility that the domain
walls will be quantum disordered.\cite{spingap2,zaanen,melted,smectic}  
In such a melted state, which might be either fully disordered (isotropic)
or still retain orientational order (``electron nematic''),
the sort  of charge and spin-ordered states that are characteristic of the
large $d$ theory occur as local correlations in the fluctuation spectrum;
a microscopic electronic theory of such quantum disordered states is not
available at present.

\section{ The Model} The model we consider is the 
straightforward generalization
of the usual $t-J$ model (or $t-J-V$ model\cite{kel}):
\ba
H= &&\frac 1 d \sum_{<i,j>} \left\{ J \vec S_i \cdot \vec S_j
+V n_i n_j\right \} \nonumber \\ - && \frac t d \sum_{<i,j>,\sigma} \left\{
c^{\dagger}_{i,\sigma}c_{j,\sigma} + {\rm H.c.} \right \}
\label{eq:tj}
\ea
where $\vec S_i=\sum_{\sigma,\sigma^{\prime}} c^{\dagger}_{i,\sigma}
\vec \sigma_{\sigma,\sigma^{\prime}}c_{i,\sigma^{\prime}}$ is the spin
of the electron on site $i$, 
$n_i = \sum_{\sigma}c^{\dagger}_{i,\sigma}c_{i,\sigma}$ is the number
of an electron on site $i$,
$c^{\dagger}_{i,\sigma}$ creates an electron
with $z$-component of spin equal to $\sigma = \pm 1/2$, $\vec \sigma$ are
the Pauli matrices, there is a constraint of no double-occupancy of
any site,
\be
n_i = 0, 1,
\ee
and $<i,j>$ signifies nearest-neighbor sites on the $d$-dimensional
hypercubic lattice.  
In comparing results of different calculations, it is important to note
that there is more than one definition of the $t-J$ model. 
Most commonly,\cite{ekl,hm}  
``the $t-J$ model'' is defined as in Eq. (\ref{eq:tj}) with $V=-J/4$, but
without the prefactor of $1/d$.  Where it can be done readily, we will quote 
results for arbitary $V$, but where this leads to complications, we will, 
for simplicity, analyze only the 
canonical case $V=-J/4$.  The additional factor of $1/d$
is included so that the ground-state energy density remains finite in the 
$d \rightarrow \infty$ limit;  thus, in making a comparison
with previous results on the $d=2$ $t-J$ model, all energies computed
here should be multiplied by  $d=2$.

\section { The undoped antiferromagnet}  The undoped system has one electron
per site so that the electron hopping term ($t$) has no effect, and the
system is manifestly insulating;  the only remaining degrees of freedom
are described by a spin 1/2 Heisenberg 
antiferromagnet with exchange coupling $J$.

\subsection{ Rigorous Bounds}  It is possible to obtain upper and lower bounds on
the ground-state energy of the spin-1/2 Heisenberg model which approach each
other in the large $d$ limit.  An upper bound is obtained by calculating 
the variational energy of the ``N\'eel'' state, which has alternating up and
down spins on alternate sites, and gives a ground-state energy per site
of $E_{Neel}= -J/4+V$.

A lower bound for the ground-state energy can be
obtained\cite{anderson} 
as follows:  We express the full Hamiltonian as a sum of pieces,
\be
H=\sum_{j=black} H_j
\ee
where the sum is over all sites on the black sublattice and $H_j$ is the 
exchange interactions between site $j$ and its nearest neighbors (which are 
necessarily on the red sublattice).  The Hamiltonians
$H_j$ are readily diagonalized, but not simultaneously since they do not 
commute with each other.  Nonetheless, the sum of 
the ground state energies of $H_j$ gives the lower bound 
$E_{lower}=-(1+d^{-1})J/4 +V$ for the ground-state energy per site.

These results, combined, prove that the ground-state energy per site, $E_{AF}$, of
the Heisenberg model approaches that of the classical N\'eel state in the limit
of infinite dimension,
\be
-(J/4) [ 1 + 1/d] \le E_{AF}-V \le -(J/4).
\ee

\subsection{ Perturbative expression for the ground-state energy and sublattice
magnetization}

We now embark on the derivation of results in a systematic expansion in
powers of $1/d$.  For this purpose, we will consider the Heisenberg model
as the isotropic limit of a Heisenberg-Ising model.  To begin with,
we use Rayleigh-Schr\"odinger perturbation theory to evaluate the
properties of interest in powers of the $XY$ coupling, and then
reorganize this perturbation theory in powers of $1/d$.  Thus, we take
as our unperturbed Hamiltonian the Ising piece of the interaction,
\be
H_0=\frac {1} d \sum_{<i,j>} \big[ J_z S_i^zS_j^z + V n_in_j\big],
\ee
and treat the $XY$ piece,
\be
H_1=\frac {J_{\perp}} d \sum_{<i,j>} [ S_i^xS_j^x+S_i^yS_j^y],
\ee
as a perturbation.  

The ground state of $H_0$ is the (two-fold degenerate) N\'eel state.  $H_1$ has
the effect of flipping pairs of spins, which because of the large coordination
in high dimensions means that the intermediate states have energies that
are proportional to $d$.  We have evaluated the perturbative
expression for the ground-state energy per site, $E_{AF}$, and the
ground-state sublattice magnetization, $m$,
to fourth order in $y\equiv J_{\perp}/J_z$, but it would be
straightforward (using modern methods of high temperature series expansion)
to extend these results to higher order. The results are:
\ba
E_{AF}= 
V-{J\over 4} \bigg[ & & 1 + {y^2 \over (2d-1) } - { y^4 (2d-3)\over 4(2d-1)^3
(4d-3) } \nonumber \\
&& + {\cal O}(y^6) \bigg ]
\ea
and
\ba 
m={1\over 2} [ 1 - y^2 {d\over (2d-1)^2} & &  +
y^4 {d[8d^3+2d-3]\over 8(2d-1)^4(4d-3)^2(d-1)} \nonumber \\
&& + {\cal O}(y^6) ].
\ea
It is clear that each power of $y$ brings with it an additional power of
$1/d$ from the additional energy denominators, 
as promised, so that the ${\cal O} (y^6)$ term is actually 
 ${\cal O} (y^6/d^5)$.
Reorganizing these expressions in powers of $1/d$ yields
\ba
E_{AF}=V-{J\over 4} \bigg [ && 1 + {y^2 \over 2d } + {y^2\over 4d^2}
+{y^2(8-y^2)\over 64d^3 } + \nonumber \\
+ && {y^2(16-3y^2)\over 256 d^4} + {\cal O}(1/d^5) \bigg ]
\ea
and
\ba
m={1\over 2} \bigg[  && 
1 -{y^2\over 4d} - {y^2\over 4d^2} - {y^2(48-y^2)\over 256 d^3} - \nonumber \\
&& {y^2(128+19y^2)\over  1024d^4} +{\cal O}(1/d^5) \bigg].
\ea
The appropriate expressions for the Heisenberg model can now be
obtained by taking the limit $y\rightarrow 1$.

\subsection{Goldstone Modes and the Long-Wavelength Physics}
Because the N\'eel state involves a broken continuous symmetry, we know that
there must exist a gapless Goldstone mode, the magnon.  In the presence of
Ising anisotropy, the magnon is massive, and is perturbatively related
to the single spin flip.  Thus, we could imagine using the same decomposition
of the Hamiltonian into an Ising and $XY$ piece to compute the magnon spectrum
perturbatively, and 
then reanalyze the expression in terms of the $1/d$ expansion.  This is impractical,
but it is instructive to see why.  

In 0$^{th}$ order ({\it i.e.} in the Ising model), there is a set of
$N/2$ degenerate excited states with excitation energy $\epsilon=J_z/2$ and
$S_z=-1$ obtained by flipping a spin on the black sublattice, and there is a
complementary set of excited states with $S_z=+1$ obtained by flipping a spin
on the red sublattice.  These states resolve themselves into the two
polarizations of the magnon band upon performing degenerate perturbation
theory in powers of $y=J_{\perp}/J_z$.  The results of degenerate perturbation
theory can be summarized in terms of an effective Hamiltonian, 
\be
H^{eff}=\sum_{i \& j = black} {\cal J}_{i,j} b_i^{\dagger}b_j
\ee
where $b_j^{\dagger}$ creates a spin flip on site $j$ and obeys boson 
commutation relations, $[b_i,b_j^{\dagger}]=\delta_{ij}$.
To be concrete, we have considered the magnon with $S_z=-1$, so
we take the Hamiltonian to operate in the
1 spin flip sector, $\sum_{j=black}  b_j^{\dagger}b_j = 1$.
The effective Hamiltonian can be solved by Fourier transform to give
a magnon energy
$\epsilon_{mag}(\vec k) = \sum_{j=black} {\cal J}_{0,j}\exp{[i\vec k \cdot \vec R_j]} $.
If we were actually interested in the case in which there was substantial
Ising anisotropy, we could simply compute ${\cal J}_{i,j}$ to the
desired order, since if $i$ and $j$ are
$n$ steps apart on the lattice, ${\cal J}_{i,j} \sim J_z [y/d]^n$, and
hence for small $y$, $H^{eff}$ is short-ranged.  It would also seem
that the same logic would justify the self-same expansion for large $d$, and
indeed (as is implicit in the discussion of the ground-state energy)
this is crudely true.  However, even though ${\cal J}_{i,j}$ falls
rapidly with $n$, the number of $n^{th}$ ``Manhattan'' neighbors grows
just as rapidly, {\it i.e.} as $ d^n$.  
For non-zero wave vector, this does not matter, as the
far neighbors contribute to $\epsilon_{mag}$ with rapidly varying phases, and
so the long-range tails of ${\cal J}_{i,j}$ are unimportant.  However,
for $\vec k$ very near $\vec k=\vec 0$ (or, equivalently, near 
$\vec k = \vec \pi \equiv
< \pi,\pi,\pi, ... >$), all terms in the Fourier transform add in phase,
so ${\cal J}_{i,j}$ must be computed to infinite order.  

Of course the point is
that the low-energy Goldstone modes have exceedingly small phase
space in large dimension, although they always dominate the
temperature dependence of thermodynamic quantities at low enough temperature 
and the asymptotic decay of correlation functions at large enough 
distances. Thus the Goldstone modes are entirely unimportant in high 
dimensions, except for physical quantities that strongly accentuate the 
lowest energy excitations.

The way to study the Goldstone behavior is in terms of
a spin-wave expansion, again suitably reinterpreted in terms of
the $1/d$ expansion. We thus start by
considering the spin-S Heisenberg antiferromagnet
in $d$ dimensions using the standard\cite{assa} 
Holstein-Primakoff bosons to obtain
the spin-wave spectrum in powers of $1/S$.
We will confine ourselves, here, to the lowest order theory, as it
adequately illustrates the point. The sublattice magnetization is
thus $S$ in the classical N\'eel state, but receives a correction of order
$S^0$ from spin-wave fluctuations as
\ba
m=S\big \{ 1 &&+{1\over 2S} \int {d\vec k \over (2\pi)^d} \big[ 1- {1
\over \sqrt{1-\gamma^2(\vec k)}}\big ]  \nonumber \\
&& + {\cal O}({1\over S^2}) \big \}
\label{eq:m}
\ea
where the integral over $\vec k$ is over the first Brillouin zone,
\be
\gamma(\vec  k) = (d)^{-1} \sum_{a=1}^d\cos[k_a]
\label{eq:gam}
\ee
is the normalized structure factor, 
and the spin-wave energy is
\be
\epsilon_{mag}(\vec k) = J S \sqrt{[1-\gamma^2(\vec k)]} [ 1 
+ {\cal O}(1/S)].
\ee

Expanding the integrand in powers of $\gamma$ and employing 
\be
\int d\vec k \mbox{ }\gamma^{2n}(\vec k)/(2\pi)^d =(2n)!/[(4d)^n n!]
\ee
we obtain 
\be 
m= S\big\{ 1- {1\over 2S}\big [ { 1\over 4d} + {\cal O}({1\over d^2})\big ]
+ {\cal O}({1\over d^2 S^2}) \big \}.
\ee
Clearly, in the limit 
$S=1/2$, the spin-wave 
expansion can be re-expressed as an expansion in powers of
$1/d$.  Indeed this expression agrees with the earlier result of perturbation 
theory in $y$, in the limit $y=1$, as, of course, it must.

For $\vec k$ near $\vec 0$ (or equivalently, near
$\vec \pi$), the spin-wave spectrum (for $S=1/2$) can be
expanded in powers of $|\vec k|$ to give the usual linear 
dispersion of the Goldstone
mode with spin-wave velocity 
\be
c=(J/d)\sqrt{d/2}[ 1 +{\cal O}(1/d) ].
\label{eq:velocity}
\ee 

We can also compute the transverse spin-spin
correlation function to leading order in $1/S$, and examine 
the resulting expression at large $d$.  Using standard results,
it is easy to see\cite{assa} that
\ba
&& <[S_i^xS_j^x+S_i^yS_j^y]> = S e^{i\vec\pi\cdot \vec R_{ij}}
 \int {d\vec k\over (2\pi)^d} \bigg \{ \nonumber \\
&& \bigg [{[1+\gamma(\vec k)]e^{i\vec\pi\cdot \vec R_{ij}}
+[1-\gamma(\vec k)]\over \sqrt{1-\gamma^2(\vec k)}} 
\bigg ]\cos{(\vec k\cdot \vec R_{ij})} \bigg \} 
\nonumber \\
&& \big [ 1 + {\cal O} ({1\over S}) \big ], 
\ea
where $\vec R_{ij}=\vec R_i-\vec R_j$.  In the large $R$
limit, this integral can be evaluated by approximating the integrand by its small $k$
expression, $1-\gamma^2(\vec k) \approx  d^{-1}k^2$, which yields the 
Goldstone behavior,
\ba
<[S_i^xS_j^x+S_i^yS_j^y]> & \sim & S\Gamma({d-1 \over 2}) {\sqrt{d}
\over \pi }
\big ({1\over \sqrt{\pi}R}\big )^{d-1}e^{i\vec\pi\cdot \vec R_{ij}} 
\nonumber \\
& \approx & {2S\over \sqrt{\pi e} } \bigg [ {R_0\over R}\bigg ]^{(d-1)}
 e^{i\vec\pi\cdot \vec R_{ij}},
\ea
where $R=|\vec R_{ij}|$, $R_0=\sqrt{d/2\pi e}$,
$\Gamma$ is the gamma function, and in the second line we have used Stirling's
formula for large $d$. It is easy to see that the integral evaluated above
is dominated by values of $k^2 \sim d^2/R^2$, and since the small
$k$ approximation is valid only so long as $k^2 \ll d$, the
Goldstone behavior is only valid for $R \gg R_0$, as is suggested by the
form of the result.

The most efficient way to evaluate properties 
of the system at low but non-zero temperature is to use the $1/d$ 
expansion to compute the fully-renormalized
zero-temperature  parameters that enter the O(3) nonlinear sigma model which
governs the Goldstone modes, namely the spin-wave stiffness, $\rho_s$, and
the transverse uniform susceptibility, $\chi_{0}$.  The susceptibility
can readily be computed  perturbatively in powers of $1/S$, and the
resulting expression reexpressed in powers of $1/d$ as
\be
\chi_0=  {1\over 4J_z}\bigg [ 1 -\frac 1 {4d}\bigg(1 + 
\frac 3 {8Sd} \bigg)+ {\cal O}({1\over Sd^{2}}) + {\cal O}({1\over 
d^2S^2}) \bigg ]
\ee 
In terms of this,  $\rho_s$  can be computed from the relation\cite{chn}
\be
\rho_s=c^2\chi_0
\ee
where $c$ is  the spin-wave velocity given in Eq. (\ref{eq:velocity}).

\section{ One hole in an antiferromagnet} 
The one-hole problem has a structure that is nominally like that
of the one magnon problem, but added factors of $1/d$ make the perturbative
approach tractable for all values of $\vec k$.  We define
as the unperturbed Hamiltonian $H_0$, the Ising limit of the $t-J$ model,
with $J_{\perp}=t=0$. 

\subsection{ The minimal hole with $S^z=1/2$}
There are $2N$ degenerate one-hole ground states of $H_0$, where the
factor of $2$ is due to the global degeneracy of the N\'eel state, 
and the factor of $N$ (which is the number of lattice sites) comes from the 
locations of the empty site (the hole).  (Henceforth we focus only on the 
states in which the magnetization is up on the black sublattice.)
These states are, in turn, separated into
disjoint Hilbert spaces labeled by the conserved quantum number, the
total $z$-component of spin,  since
a hole on a black sublattice site has $S^z=+1/2$ and one on the red sublattice
has $S^z=-1/2$.  For concreteness, we will focus on the $N/2$ degenerate
states corresponding to a hole on the black sublattice.

We now use degenerate perturbation theory
to construct the effective Hamiltonian of one hole,
\be
H^{eff}_1=\sum_{i\& j=black} t_{ij}c_i^{\dagger}c_j+ 
\sum_{i\& j=red} t_{ij}c_i^{\dagger}c_j,
\label{eq:heff1}
\ee
where $c_j^{\dagger}$ is the fermionic creation operator for a hole 
on site $j$, and
for Hermiticity, $t_{ij}=t_{ji}$.  Once the effective Hamiltonian 
is computed, its
eigenstates and eigenvalues can be found by Fourier transform:
\be
\epsilon_{hole}(\vec k)=\sum_{j=black}t_{0j}\exp[i\vec k\cdot \vec R_j].
\label{eq:ephole}
\ee

To begin with, we study the perturbative expressions for
the diagonal term, $\epsilon_0\equiv t_{ii}$.

\ba
\epsilon_0= -2V+{J_z\over 2} && \bigg[ 1 + y^2{2(d-1)\over (2d-1)(4d-3)} - 
z^2{8\over (2d-1)} \nonumber
\\ && + 
{\cal O}(y^4) +{\cal O}(z^4) +{\cal O}(y^2z^2)\bigg]
\\
=  -2V+{J_z\over 2} && \bigg[ 1 + {[y^2 - 16z^2]\over 4d} + 
{[y^2 - 32z^2]\over 16d^2}
+{\cal O}(1/d^3)\bigg] \nonumber
\ea
where $z=t/J_z$ and $y=J_{\perp}/J_z$.  The corrections to the hole self energy
which are independent of $y$ are the famous string corrections to the hole
self energy, of which the retraceable paths of Brinkman and Rice\cite{brinkman} 
are a subclass.

Next we study the coupling between second ``Manhattan''
neighbor sites (nearest neighbors on
the black sublattice).  There are $2d(d-1)$ ``true second neighbors'', 
reached by taking
a step to the nearest neighbor site in one direction, and then
a second step in an orthogonal direction, and there are $2d$
``straight line'' second neighbors reached by taking two 
steps in the
same direction.  For $i$ and $j$ true second neighbors, 
$t_{ij} \equiv 2\tau+\tau_{Ising}$, 
where the factor of $2$ in the definition takes account of the fact that 
there are two minimal paths to the true second
neighbor. Here $\tau$ is given by
\ba
\tau =&&{4tyz\over d(2d-1)(4d-3)} \times \nonumber \\
&& \ \  \bigg[ 1 - y{(6d^3-6d^2+1)\over 3(d-1)(2d-1)(4d-3)}
-y^2{N_2\over
D_2} \nonumber \\
&& \ \ -z^2
{(528d^4-1368d^3+1160d^2-314d-7)\over 3(2d-1)^2(d-1)(4d-3)(6d-5)}\nonumber \\
&& + {\cal O}(y^3) + {\cal O}(z^4) + {\cal O}(z^2y) \bigg]
\nonumber \\
= && {tyz\over 2d^3} \bigg[ \big(1-\frac y 4\big)
+\frac 5 {4d}\big(1-\frac y 2- \frac {22z^2} {15}\big) +
{\cal O}(1/d^2) \bigg],
\label{eq:t2}
\ea
with 
\ba
N_2 = && 294912d^7-1834752d^6+5070688d^5 - 7924688d^4 \nonumber \\
&&  +7461296d^3 - 4195852d^2 + 1298748d - 170325
\ea
and 
\ba
D_2= && 96(2d-1)^2(d-1)(3d-2)(4d-3)^2  \nonumber \\
&& \ \ \times (6d-5)(8d-9)(8d-7);  
\ea
$\tau$
comes from processes in which, in lowest order, the hole hops twice, followed 
by a spin exchange which repairs the resulting damage to the spin order.
In addition, there is a contribution
\ba
\tau_{Ising}= && -{32tz^5\over 3d(2d-1)^2(4d-3)^2(d-1)}\bigg[ 
1 + {\cal O}(z^2) \bigg] \nonumber \\
= && -{tz^5\over 6d^6}\bigg [
  1 + {\cal O}(\frac 1 d) \bigg].
\ea
which comes from a process in which a hole circles a plaquette one and a 
half times, thus ``eating its own string''. This process, which was discovered 
by Trugman\cite{trugman} for $d = 2$, survives even in the Ising limit $y=0$.
However, $\tau_{Ising}$ is higher order in powers of $1/d$ and so is 
negligible, even when $t \gg J$.
For straight-line second neighbors, $t_{ij}=\tau^{\prime} +
\tau^{\prime}_{Ising}$
where 
\ba
\tau^{\prime}=&&{4tyz\over d(2d-1)(4d-3)} \bigg[ \nonumber \\
&&  1 - y{(d-1)\over (4d-3)} +
y^{2}\frac{N_{2}^{\prime}}{D_{2}^{\prime}} + 
z^{2}\frac{N_{3}^{\prime}}{D_{3}^{
\prime}} \nonumber \\
&& {\cal O}(y^3) + {\cal O}(z^4) + {\cal O}(z^2y) \bigg]
\nonumber \\
= && {tyz\over 2d^3} \bigg[ \big(1- \frac y 4 \big)
+\frac 5 {4d}\big(1-\frac {y} {5}-\frac {22z^2} {15} \big) +
{\cal O}(1/d^2) \bigg],
\label{eq:tauprime}
\ea
where
\ba
N_{2}^{\prime} = && 73728 d^{6} -41088 d^{5} -514544 d^{4}
 +1145088 d^{3}\nonumber \\
&& \ \   -1016784 d^{2}+421204 d -67611 
\ea
\ba
D_{2}^{\prime} = && 48(2d-1)^{3} (3d-2)(4d-3)^{3}(6d-5)  \nonumber \\
&& \ \  \times (8d-9) (8d-7)
\ea
and
\ba
N_{3}^{\prime} = -176 d^{3} +280 d^{2} -108 d -2 
\ea
\ba
D_{3}^{\prime} = (2d-1)^{3}(4d -3)^{2} (6d-5)
\ea
and $\tau^{\prime}_{Ising}$ is the corresponding Trugman term    
which is ${\cal{O}}(tz^{9}/d^{9})$. 
Because they are higher order in $1/d$, we will henceforth
ignore $\tau_{Ising}$ and $\tau^{\prime}_{Ising}$ relative to $\tau$.
However, we shall see below that the difference, 
\ba
\tau^{\prime}-\tau =&& {4tzy^2(3d-2)^2\over 3d(d-1)(2d-1)^2(4d-3)^2}\bigg[ 
1 + {\cal O}(y) +{\cal O}(z^2)\bigg] \nonumber \\
=&& {3tzy^2\over 16d^4}\bigg[ 1 + {(13+y)\over 6d} + {\cal O} (d^{-2}) \bigg],
\label{eq:dtau}
\ea
although
smaller than $\tau$ by a factor of $1/d$, plays a
critical role in determining the band structure of the minimum energy
hole in an antiferromagnet.

The sign of these terms deserves some
comment.  Since the lattice structure defined by $t_{ij}$ is not
bipartite, the sign of the matrix elements is physically significant.
In the present case, since the leading order contributions to $\tau$ come
from third order perturbation theory, the resulting matrix element is 
positive.  

These expressions may be combined to obtain an expansion for the bandwidth 
$W$ of a single hole by adding the contributions from the different hopping 
processes. Each hop contributes a factor 2$\tau$ or 2$\tau^{\prime}$ to $W$, so
\be
W = 4d(d-1)\tau + 4d \tau^{\prime}. 
\ee
Then, using Eqs. (\ref{eq:t2}) and (\ref{eq:tauprime}) and expanding in powers of
$d^{-1}$,
\be
{dW \over t} = 2yz(1-{y \over 4}) + {5yz \over 2d}(1-{y \over 2} -{22z^2 \over 15}) 
+ {\cal O}(d^{-2}).
\label{eq:bandwidth}
\ee
This result does not make sense unless $W>0$ or
\be
z^2 < {15 \over 22} \big[{4d \over 5}(1- {y \over 4}) + 1 - {y \over 2} \big ]
\label{eq:wpos}
\ee
Since $z=t/J_z$, this illustrates the fact that the motion of a single hole
is exchange dominated in the large-$d$ expansion.

It is important to note that the leading order expression for $\tau$ 
contains one more power of $1/d$ than the 
corresponding matrix element in the effective Hamiltonian for one magnon.
Indeed, the leading-order behavior of the matrix elements connecting 
sites separated
by $2n$ nearest-neighbor steps ($2n^{th}$ nearest Manhattan neighbors) is
$t_{2n} \sim tz^{2n-1}y^n/d^{3n}$, where the factor of $tz^{2n-1}\propto
 t^{2n}$ comes from
the minimum number of hops for the hole to 
propagate this distance, the factor of $y^n\propto J_{\perp}^n$ 
reflects the minimum number of spin flips needed to restore the spins to
a ground-state configuration
following the passage of a hole, the factor of $d^{-(3n-1)}$ comes from the
accompanying energy denominators at this order of perturbation theory, and one
additional factor of $1/d$ comes from the overall normalization of
the Hamiltonian.  Because of these
extra factors in the expression for $t_{ij}$, its contributions to 
the hole energy
fall rapidly with distance in high dimensions, where the number of 
$2n$-step paths
is $2d(2d-1)^{2n-1}$, and so the number of $2n^{th}$ Manhattan neighbors can grow no
faster that $(2d)^{2n}$.  Thus, the longer range pieces of $t_{ij}$ can be neglected
for any value of $\vec k$ in large enough dimension.

The effective Hamiltonian obtained by retaining only terms out to second 
Manhattan neighbors is given by 
\ba
\epsilon_{hole}(\vec k) = &&\epsilon_0-2d(\tau^{\prime}-\tau)
+ (2d)^2\tau\gamma^2(\vec  k) \nonumber \\ + 
&& 4(\tau^{\prime}-\tau)  \sum_{a=1}^d\cos^2[k_a].
\label{eq:hole}
\ea
If we ignore the small difference, $(\tau^{\prime}-\tau)$, then the minimum energy
hole states are located on the $d-1$ dimensional hypersurface, $\gamma(\vec k)=0$;
since the band-structure for the noninteracting tight-binding model is
$\epsilon_{free}=-2t\gamma(\vec k)$,  and that model is, in turn,
particle-hole symmetric, this hypersurface is  precisely the  Fermi surface of
the half-filled band in the absence of interactions.  With higher
order terms in powers of $1/d$ (which produce a non-zero value of
$(\tau^{\prime}-\tau) > 0$) the minimum energy of a single hole occurs at 
$\vec k = (1/2)\vec \pi$ and the $2^d-1$ symmetry-related points.

We emphasize that, although we have treated
the effects of $t$ perturbatively, we have not made a small $t$ approximation.
The present results are valid for arbitrary $t/J$, so long
as it is not parametrically large ({\it i.e.} so long as $t\ll dJ$).
Nonetheless, because each hop of the hole may break ${\cal O}(d)$ bonds,
the large-$d$ limit is exchange dominated.

\subsection{Magnetic polarons with larger spin}
In low dimensions, and for $t\gg J$, it is believed that a single hole in an antiferromagnet
produces a ferromagnetic bubble in its vicinity, or, more
precisely, that there is
a series of level crossings as a function of $t/J$ 
at which the total $z$ component $S^z$ of the spin of the ground state
for a single hole state steadily increases.  However, in the
large $d$ limit, the antiferromagnetic energy always dominates unless
$t$ is parametrically larger than $J$, {\it i.e.} unless $t\sim d^x J$, 
where 
$x$ is a positive exponent (which we will estimate below).  Such parametrically
large values are beyond the scope of the present analysis.

To estimate the magnitude of $t/J$ at which ferromagnetic bubbles first appear, we
consider
a straightforward generalization (to $d > 2$) of the calculation of 
Emery, Kivelson, and Lin\cite{ekl} for a hyperspherical
ferromagnetic polaron in the large 
size limit (where discrete lattice effects can be neglected).  We balance the magnetic
energy lost in the volume of the polaron against the zero-point energy (computed in
the effective mass approximation) to localize the hole in the interior of the polaron.
This results in a polaron with a radius, 
$L\approx [2t\pi/A_d(V-E_{AF}+J/4)]^{1/(d+2)}$,
where $A_d$ is the area of the unit $d$ dimensional hypersphere,
\be
A_d=2(\sqrt{\pi})^d/\Gamma(d/2)\approx \sqrt{d/\pi}(\sqrt{2\pi e/d})^d,
\label{eq:Ad}
\ee
and in the final line, we have used Stirling's approximation for large $d$;
the polaron has spin $S^z \approx  A_d L^d/d$.  By definition, the spin of
the polaron must be substantially greater than 1, which means that $t \gg d^2 J$. 
We conclude that in the large $d$ limit, the low energy hole branch is always
the naive, $S^z=1/2$ vacancy state, and that local ferromagnetism is never
a relevant piece of the one hole physics.

\section{Effective interactions between two holes}

Broadly speaking, the effective 
interactions between two holes are of two kinds -- ``potential'', 
which are induced by distortions in the antiferromagnetic order, and 
``dynamic'', which minimize the zero-point kinetic energy of a hole.

At long distances, the effective potential of interaction can be computed 
by considering change in the magnetic Hamiltonian induced by two
static holes at lattice sites $0$ and $i$,
\be
H_2=-{J\over d}\bigg[\sum_{j}^{(i)}\vec S_i \cdot \vec S_j
+ \sum_{k}^{(0)} \vec S_0 \cdot \vec S_k\bigg]
\ee
where $\sum_{j}^{(i)}$ signifies the sum over the nearest neighbor sites of 
$i$. Then, on integrating out the magnetic degrees of freedom, we obtain
\ba
&&V^{eff}(\vec R_i)= - {J^2/d} \int dt\sum_{j}^{(i)}\sum_{k}^{(0)} \nonumber\\  
&&\bigg <T\big[\vec S_i(t)\cdot \vec S_j(t) -<\vec S_i \cdot \vec S_j>\big]
\big[\vec S_0 \cdot \vec S_k -<\vec S_0\vec \cdot S_k>\big] \bigg > \nonumber \\
&& + ... 
\ea
where $ ... $ are higher order  terms in powers of $J$,
which are also of shorter range as a function of $|\vec R_i|$, $t$ 
is imaginary-time, and $T$ is the imaginary time ordering
operator.  It is straightforward\cite{leonid} 
to determine from linear spin-wave theory
that 
\be
V^{eff} \sim 1/R^{(2d-1)}
\ee
which is a short range potential in the sense that the integral
over space of $|V^{eff}|$ is non-infinite in all dimensions greater than $d=1$.
(The integral over space of $V^{eff}$ itself is easily seen to be 0.)
Moreover, the long distance tails of $V^{eff}$ decrease in importance as 
$d$ increases.
For this reason, we will ignore the power-law tails of $V^{eff}$ and
simply consider its dominant, short-distance pieces.

The nearest-neighbor interaction between two holes (which is certainly
attractive in the canonical case, $V=-J/4$) can readily be
computed from perturbation theory in powers of $y=J_{\perp}/J_z$:
\ba
V^{eff}(\hat e_1)=&& {V \over d}
-{J_z\over 4d}\big[ 1  + {\cal O}
(y^4) \big] \nonumber \\
=&& {V\over d}-{J_z\over 4d}\big[ 1 
+{\cal O}(1/d^3)\big]. 
\label{eq:v2hole}
\ea
There is a considerably weaker interaction (which, in fact, is
repulsive) between second 
nearest-neighbor holes:
\ba
V^{eff}(\hat e_1+\hat e_2)=&&{J_z y^2\over 4d(2d-1)(4d-3)}\big [
1 + {\cal O}(y^2)\big ] \nonumber \\
=&& {J_z y^2\over 32 d^3} \big[ 1 + {\cal O}(1/d) \big].
\ea
Indeed, to this order, the effective interaction is the same for all second
Manhattan neighbors, $V^{eff}(2\hat e_1)= V^{eff}(\hat e_1+\hat e_2)[1
+{\cal O}(1/d^3)]$.
Clearly, for further Manhattan
neighbors, the effective interactions are down by additional powers of $1/d$.

The kinetic terms, in general, generate fairly complicated interactions of the
form
\be
T^{eff}=\sum_{ijkl}T_{ijkl}c_i^{\dagger}c_j^{\dagger}c_kc_l
\ee
where, as before, $c_i^{\dagger}$ is a fermionic creation operator
for a hole at site $j$.  However, in large $d$, it is strongly dominated
by its short-range components, of which the dominant terms are a
potential interaction between 
nearest-neighbor holes (which renormalizes $V^{eff}(\hat e_1)$) and
a pair-hopping term.  Indeed, combining the potential and kinetic
terms to leading order in $1/d$, we find a two hole contribution
to the effective Hamiltonian ({\it i.e.} the interaction part of the
effective Hamiltonian, of which Eq.  (\ref{eq:heff1}) is the
noninteracting piece):
\ba
H^{eff}_2=&& U^{eff}\sum_{<i,j>}c_i^{\dagger}c_j^{\dagger}c_jc_i  \\
\label{eq:heff2}
&& -T^{eff}\sum_{<i,j,k>}c_j^{\dagger}c_i^{\dagger}c_jc_k + {\cal O}(1/d^3)
\nonumber
\ea
where in the pair-hopping term
$<i,j,k>$ signifies a set of sites such that $i$ and $k$ are both
nearest-neighbors of $j$ (which we define to include the case $i=k$), and
\ba
U^{eff}=&&{V\over d} -{J_z\over 4d}\big[ 1 + 
{8z^2\over (d-1)(2d-1)} \nonumber \\
&& + {\cal O}(y^4) + {\cal O}(z^4) + {\cal O}(y^2z^2)\big] \nonumber \\
=&&{V\over d} -{J_z\over 4d}\big[ 1 +  {4z^2\over d^2}
+{\cal O}(1/d^3)\big]. 
\ea
and
\ba
T^{eff}=&& {tz\over d (d-1)}\big[ 1 + {y\over 4(d-1)} + {\cal O}(y^2) 
+ {\cal O}(z^2)\big] 
\nonumber \\
= && {tz\over d^2}\big [ 1 + {1\over d}(1 + {y\over 4}) 
+ {\cal O}(1/d^2) \big ].
\ea

The pair-hopping term, $T^{eff}$, has an interesting history:
In early work on high temperature superconductivity, it was often claimed
that, whereas the motion of a single hole is inhibited by antiferromagnetic 
order, pair motion appears to be entirely unfrustrated. It was suggested that
this might indicate a novel (non-potential) source of an attraction between 
holes which could be the mechanism of high temperature superconductivity.  
At first sight, the fact that $T^{eff}\sim
d^{-2}$, while the single-particle hopping term is $\tau\sim d^{-3}$,
appears to support the validity of this idea in large $d$.
However, the fallacy of this argument was
revealed in the work of Trugman\cite{trugman}, who showed that
this mode of propagation of the hole pair was frustrated by
a quantum effect which originates in the fermionic character of the 
hole.  In large $d$, this frustration effect is particularly graphic.
Pair binding is enhanced if we ignore single-particle hopping, and 
diagonalize $H^{eff}_2$.  Of course, any state in which the two holes are 
farther
than one lattice site apart are eigenstates of $H_2^{eff}$, so long as
terms of this range are neglected (because they are of higher order) as in
Eq. (\ref{eq:heff2}).  For 
the states in which the two holes are nearest-neighbors, $H_2^{eff}$ can
be block diagonalized by Fourier transform, with the result that there
are $d$ bands of eigenstates labeled by a band index and a
Bloch wave-vector, $\vec k$.  It is straightforward to see that none
of these bands disperses (their energies are independent of $\vec k$)
and that $d-1$ of these bands have energy $U^{eff}$, while
the remaining band has energy $ U^{eff}+2T^{eff}$.  This final band,
which feels the effect of pair propagation, has the highest energy.  On the
other hand, if the holes were bosons, this latter band
would have energy $U^{eff}-2T^{eff}$,
which is much closer to what one might have expected.

It follows from this argument that coherent propagation of a pair is not an 
effective mechanism of pair binding and that the short-range attraction 
between two holes in an antiferromagnet arises from the fact that
two, nearest-neighbor holes break one less antiferromagnetic bond than
two far-separated holes. This interaction is sufficient to produce a two-hole
bound state because the one-hole spectrum has an essentially
degenerate band minimum along the $d-1$ dimensional magnetic Brillouin zone.
As in the Cooper problem, this gives a constant density of states at low
energy and any attractive interaction is sufficient to produce a bound state.

\section{Finite hole concentration}  

We have suggested \cite{ekl,losalamos}
that, in general, a doped antiferromagnet in $d\ge 2$ will phase separate 
into a hole-free antiferromagnetic region and a hole-rich region.
There is now substantial evidence, both numerical\cite{ekl,hm} and
analytical\cite{ekl,assaS,assaN}, that this is the case for 
the $t-J$ model in $d=2$.  Phase separation is, of course, a first order
transition, so it must be studied by comparing the total energy of various
candidate homogeneous and inhomgeneous states to find the true ground state
at fixed hole density.  

In large dimension, there are many metastable states 
which are, in a sense, ``local'' ground states of given character.  While
the large $d$ limit allows us to compute the energy and character of
a given candidate ground state exactly, it is almost never possible
to prove that we have actually identified the global ground state.
Specifically, we have computed the energy of various candidate states 
(as discussed below) and found that, of these, the lowest-energy
state is phase separated into an undoped antiferromagnet
and a hole-rich phase with a very low {\it electron} density,
{\it i.e.} with hole concentration equal to or nearly equal to 1.
Moreover, given that the interaction between two holes
is strongly attractive (of order the hole bandwidth) at short distances,
and weakly attractive at long distances, we feel that 
it is extremely unlikely that any dilute hole-liquid or hole-crystal phase in
the antiferromagnet is stable in large $d$.  Below, we show that the
same instability shows up in
a dilute domain-wall phase in which the holes are concentrated on an array
of widely separated domain walls.  

What this means is that, under the assumption that we have not overlooked
a lower-energy state (which we consider unlikely), we
can obtain a complete and exact understanding of the zero-temperature
phase diagram in the $d\rightarrow \infty$ limit by considering
the phase coexistence between the undoped antiferromagnet (which we have
already characterized) and a
very hole rich or dilute electron phase.

We emphasise that the physics of phase separation 
and domain walls (Sec. VII) in large-$d$
is not exchange dominated becuse it does not involve 
breaking a large number of bonds.

\subsection{Properties of dilute electrons in large $d$ }

We now consider the ground-state properties of dilute electrons on a 
$d$-dimensional hypercubic lattice, for $d$ large.
Near the bottom of the band, the electron dispersion relation is approximately
quadratic, {\it i.e.} $\epsilon(\vec k)\approx
-2t + tk^2/d$ is a good approximation so long as each component of
$\vec k$ is small compared to $1$, or typically, that $k^2 \ll d$.  Also in this
limit, the interactions between electrons are weak (since they rarely approach
each other), and so can be ignored to first approximation.  Thus, we will begin
by considering the properties of the non-interacting, quadratically
dispersing electron gas in $d$
dimensions.  For this problem, the Fermi momentum as
a function of the
chemical potential, $\mu$, is $k_F=\sqrt{(\mu+2t)d/t}$ for
$\mu>-2t$, and the corresponding density is 
\be
n={2A_d\over d} \bigg({k_F\over 2\pi}\bigg)^d = {2\over\sqrt{\pi d}}
\bigg(\sqrt{e\over
2\pi d} k_F\bigg)^d\big [ 1 + {\cal O}({1\over d})\big ]
\ee
where $A_d$ is given in Eq. (\ref{eq:Ad}),  a spin-degeneracy factor of $2$ has
been included, and the final equality uses the large $d$ expression
for $A_d$.  Note that whenever the condition $k_F^2 \ll d$ is satisfied, the
electron density is exponentially small for large $d$, so our approximations
are exponentially accurate.  
The energy per site of this system can be computed readily:
\be
E_{gas}= -2tn[ 1 - k_F^2/(2d+4) ].
\ee

\subsection{Conditions of thermodynamic equilibrium}

In general,   for two phases to be in thermodynamic
equilibrium, they must have equal chemical potentials, $\mu$.  
However, here, 
the undoped antiferromagnetic phase is incompressible,
so that the zero temperature chemical potential is undetermined. Then the
condition for the electron gas to be in equilibrium with
 the antiferromagnet  is
\be
\mu=[E_{AF}-E_{gas}(\mu)]/[1-n(\mu)]
\label{eq:mu}
\ee
where $E_{gas}$ is the ground-state energy per site of the electron gas and
$n$ is the electron density, both of which are funtions of $\mu$.
Thus, if $E_{AF} < -2t$,
the only thermodynamically  stable zero temperature
phases of the $t-J$ model are the undoped antiferromagnet and the vacuum (no
electrons).  On the other hand, if $E_{AF} > -2t$, phase coexistence is
possible between the antiferromagnet and a ``metallic'' phase with
allowed electron densities, $n\le n_{max}$, where
$n_{max}$ is the electron density at  the equilibrium value of $\mu$.
We shall see shortly that both $E_{gas}$ and $n$ are exponentially small
at large $d$, so that to exponential accuracy, 
\be
\mu=E_{AF}.
\ee
If we
use the large $d$ expression for $E_{AF}$, we conclude that the metallic
state is stable only if $J < 4t[ 1 + {\cal O}(1/d) ]$, and that 
if this condition
is satisfied, the maximum
stable density of the metallic phase is
\be
n_{max}={2\over \sqrt{\pi d}}\bigg[\sqrt{e(4t-J)\over 4\pi t}
\bigg]^d\big [ 1 +
{\cal O}({1\over d}) \big].
\label{eq:nmax}
\ee
Notice that this quantity is small (and hence our approximations are
justified), even in the limit $t \gg J$, where
\be
n_{max} \rightarrow (2/ \sqrt{\pi d}) \big[{e\over \pi}\big]^{d/2}.
\label{eq:nmaxmax}
\ee

\subsection{Effective Interactions in the Metallic State, and the
Conditions for Superconductivity}

Since the electron density in the metallic state is small, interaction
effects are dominated by pairwise collisions between electrons.  In the
triplet channel, there is a nearest-neighbor electron-electron repulsion
of strength $(4V+J)/4d$, while in the singlet channel there is an infinite,
on-site repulsion, and a nearest-neighbor attraction of strength $(4V-3J)/4d$.
For the canonical choice of $V=-J/4$, which we adopt in most of this section
for simplicity of notation, the attractive interaction in the singlet channel 
is simply $-J/d$.

We can look for evidence of a simple s-wave instability of the
metallic state in the low density limit
by studying the conditions for the existence
of a solution to the BCS gap equation.  
First, consider the unperturbed (noninteracting)
thermal Green function at low, but finite temperature
\be
G({\vec k})={\tanh\big[\frac {\beta} 2
(\epsilon(\vec k)-\mu)\big ] \over 2[\epsilon(\vec k)-\mu]}
\ee
where $\epsilon(\vec k) = -2t \gamma(\vec k)$ and $\gamma(\vec k)$ is defined in
Eq. (\ref{eq:gam}). Now it is straightforward to show that, 
in the singlet channel, the BCS equation for
the transition temperature T$_c$ may be written in terms of
the corresponding real-space Green functions 
\be
G_0\equiv \frac 1 N \sum_{\vec k} G(\vec k),
\ee
\ba
G_1\equiv &&\frac 2 N \sum_{\vec k} G(\vec k)\sum_{a=1}^d \cos(k_a) \nonumber \\
=&& -\frac d {tN}\sum_{\vec k} \epsilon(\vec k)G(\vec k),
\ea
and 
\ba
G_2\equiv &&\frac 2 {dN} \sum_{\vec k} G(\vec k)[\sum_{a=1}^d \cos(k_a)]^2. \nonumber \\
= &&\frac d {2t^2N} \sum_{\vec k} \epsilon^2(\vec k)G(\vec k)
\ea
as the $U \rightarrow \infty$ limit of
\be
0=(1+UG_0)(1-JG_2/d)+JUG_1^2/2d^2,
\ee
or, in other words,
\be
\frac 1 J = \frac {G_2} d - \frac {G_1^2} {2d^2G_0}.
\label{eq:bcs}
\ee

Since $\epsilon(\vec k)$ appears in the numerators of $G_1$ and $G_2$, as
well as in the denominator of $G(\vec k)$, 
one may express $G_1$ and $G_2$ in terms of $G_0$
and sums in which $\epsilon(\vec k)$ does not appear in the denominator.
Now, to find the critical ratio of $t/J$ at which $T_c\rightarrow 0$, 
note that, for $\mu$ near the bottom of the band ($\mu\approx -2t$),
the approximation $\tanh[\beta(\epsilon(\vec k)-\mu)/2] \approx 1$
can be used in the various sums, except in $G_0$. Then Eq. (\ref{eq:bcs}) becomes
\be
\frac 1 J - \frac 1 {2t} =  - \frac 1 {8t^2 G_0}
\label{eq:cond}
\ee 
Consequently the effective interaction is attractive and the BCS 
transition temperature for s-wave pairing is non-zero, so long as
\be
J/t > 2.
\ee
This result is valid for dilute electrons in any dimension greater than 1,
and is in agreement with earlier results in two dimensions.\cite{ekl,lin,hm2}
Notice that Eq. (\ref{eq:cond}) with $\tanh[\beta(\epsilon(\vec k) -\mu)/2]$
set equal to 1 in $G_0$ is the condition for
a two-hole bound state, but in this case 
$G_0$ is divergent only in $d=1$ and $d=2$.
Thus, in higher dimensions, the critical value of $J/t$ for a two-hole
bound state depends on $d$, and tends to infinity as $d\rightarrow \infty$.

For $J/t<2$, the direct pair interaction is repulsive in all channels,
so there is no solution of the BCS equations.  However superconductivity 
could emerge in a higher angular-momentum channel by a variant of the 
Kohn-Luttinger effect. This sort of
instability has been studied extensively in two dimensions,\cite {hm2} and 
could probably be analyzed in much the same way here.  However, 
as the electron density
is anyway exponentially small in large $d$, and these effects are of still
higher order in the density, they occur at energy scales that are 
exponentially smaller than the exponentially small Fermi energy.

One can, of course carry through the same analysis for arbitrary values
of $V$, in which case, Eq. (\ref{eq:cond}) is replaced by
\be
{3J \over 4} - V > 2t
\ee
and, in general, there is no superconducting transition if $V > 3J/4 -2t$.
Recently, Riera and Dagotto\cite{riera} have shown that a sufficiently 
large value of $V$ prevents bound states of pairs of holes
in the two-dimensional $t-J$ model.

Because there are no particularly good nesting vectors, we consider it
unlikely that the hole-rich phase is subject to any sort of charge
or spin-density wave instability of the hyperspherical Fermi surface in 
large dimension. 

\section{One Domain wall}  
As mentioned previously,
one feature of this large-$d$ perturbation theory 
is that there are many interesting states which, if not the global
ground state of the system, are at least the lowest-energy eigenstates 
in a restricted sector of the Hilbert space.  
In particlar, domain-wall states are interesting in their own right,
although they prove to be of special 
importance in the presence of long-range Coulomb interactions.

First of all, we define a domain wall to be a $d-1$ dimensional
hypersurface which cuts the full lattice
in two, such that far from the domain wall the system is undoped, and hence
antiferromagnetically ordered.  Since, as we shall see, in large $d$ such domain
walls are always smooth (transverse quantum fluctuations of the position of 
the domain wall
are suppressed by powers of $1/d$), we can characterize such a state
by the mean position of  the domain wall, its net charge ({\it i.e.} the
concentration of holes per hyperarea) and the phase shift (if any) suffered 
by the antiferromagnetic order in crossing the domain wall.  In addition,
since domain walls lie along lattice symmetry directions 
in all cases of interest,
we can classify them by the broken crystal point group symmetry.
Here we analyze what we
believe to be the physically most important domain walls, but we have not 
yet carried out an
exhaustive study of other possible domain wall structures.

\subsection{``Vertical'', site-centered domain wall}  A vertical domain wall
lies perpendicular to a principal axis of the hypercube, 
which we call the ``x-axis'',
and its location is thus specified by a single $x$ coordinate.
If the domain wall preserves reflection symmetry relative to a 
hyperplane that is perpendicular
to the x axis and crosses the x-axis at a lattice  site (which we will take to
be $x=0$), the domain wall is said to be site-centered;  if  this reflection
plane passes half-way between two adjacent lattice sites 
(which we will take to be $x=0$ and
$x=1$), the domain wall is said to be bond-centered.  We begin with
the case of a vertical, site-centered domain wall with charge density of
one hole per site, {\ie} one that corresponds to a hypersurface of empty sites.

We thus consider as the unperturbed part of the Hamiltonian, $H_0$
the Ising part of
the $t-J$ model with $t=0$, but possibly with different 
choices of the Ising $z$-axis to the right ($x>0$) and to the
left ($x<0$) of the domain wall.  Then $H_1$ is all the remaining interactions,
including all terms involving $t$.  

The ground state of $H_0$ in the appropriate charge sector is four-fold
degenerate:  it has no electrons
in the domain wall (one hole per site) and one of the two possible N\'eel 
states in each of the two ``halves'' of the system.  The energy per
domain-wall site (relative to the energy of the uniform, undoped N\'eel state) can
now be calculated perturbatively, and it can be seen straightforwardly that
the perturbation theory is again controlled by large $d$;  for instance,
from second order
perturbation theory we find
\ba
\epsilon_{vert}=&& {(d+1)J_z\over 2d} \bigg\{ 1 + {y^2(4d^2 -3d-2)\over (d+1)(2d-1)(4d-3)}
\bigg\}\nonumber\\
&& -{8J_zz^2 
\over [2d-1+\cos(\delta)]} + ... \nonumber \\
=&& {J_z\over 4} \bigg\{ 1 +{(2 + y^2)\over 2d} +{(y^2 -64z^2)\over 4d^2}
+{\cal O}(1/d^3)\bigg\}, \nonumber
\ea 
where $\delta$ is the phase shift of the N\'eel order across the domain wall.

It is clear that 
to this order in $1/d$, the energy is independent of $\delta$.  However, from the
perturbative expression, we can extract the leading order $\delta$-dependence:
\be
\epsilon_{vert}(\delta)-\epsilon_{vert}(\pi)={2 t^2\over J_z} 
{[\cos(\delta)+1]\over d^2}
\bigg\{ 1 + {\cal O}(1/d)\bigg\}. \nonumber
\ee
Clearly, then, the energy is minimized by an ``antiphase'' domain wall, with
$\delta=\pi$. The preference for an antiphase domain wall is produced by the 
transverse fluctuations of the  domain wall, and it is likely to survive in 
all lower dimensions. The only cost of large transverse amplitude fluctuations 
of an antiphase domain wall is an increase in the surface energy, whereas, for 
any other $\delta$, there are regions of impaired spin correlations.  Indeed, 
we know from the solution of the one-dimensional electron gas with repulsive 
interactions that, near half filling, the holes are solitons that act locally 
like antiphase domain walls in that the magnetic correlations are
shifted by $\pi$ as one passes a hole, although of course there is no long
range magnetic order in this case.

The empty domain wall we have been considering is locally stable so long
as $J/t$ is sufficiently large ($J/t > Y_c = 4[ 1 + {\cal O}(1/d)$]~),
but is unstable for smaller values of $J/t$.  To see this, consider 
the state in which one electron is transferred from the surrounding
antiferromagnet to the bottom of the domain-wall band.  In first order
degenerate perturbation theory in $t$, the bottom of the domain wall
band is easily seen to be $\epsilon=-2t(d-1)/d + ...$, {\it i.e.} equal
to the band bottom of the dilute electron phase to leading order in $1/d$.
This means that, for the domain wall to be in local equilibrium
with the surrounding antiferromagnet, it must have approximately 
(up to higher order corrections in $1/d$) the
same electron concentration, $n_{max}$ in Eq. (\ref{eq:nmax}), as the
hole-rich phase which can exist in equilibrium with the antiferromagnet.
It is in this sense that a domain wall can be viewed as a form of
local phase separation.  Since even when $J/t << 1$, $n_{max}$ is
exponentially small, the existence of dilute electrons within the
domain wall when $J/t < Y_c$ does not significantly affect any of the
other calculations described above.  However it does imply that these 
electrons render the domain wall metallic and, under appropriate
circumstances, superconducting.  This, of course, implies a qualitative
difference in the electronic properties of domain walls for small and
large $J/t$.

\subsection{Bond-centered vertical domain wall}

If we fix the hole density at one (or approximately one) hole per
hypersurface unit cell, as above, it is immediately clear to $0^{th}$
order in $1/d$ that a bond-centered
domain wall will have considerably higher surface tension than a
site-centered wall:  On the one hand, 
more antiferromagnetic bonds are broken by
the bond-centered wall.  
On the other hand, only a very small fraction
of the states in the electronic band have energy of order $-td^0$, so
even if we ignored the constraint of no double occupancy, and allowed
the remaining electrons in the domain wall to minimize the kinetic
energy part of the Hamiltonian, the gain in kinetic energy will be higher order
in $1/d$ than the loss of exchange energy.

Of course, if we were to roughly double the concentration of holes, then
a bond-centered domain wall can be viewed as simply two nearest-neighbor
site-centered domain walls.  This situation will be considered, below,
when we consider interactions between domain walls.

\subsection{Other Domain Walls, Domain-wall Kinks, etc.}

There are many other kinds of domain walls.  For instance,
one could consider a ``diagonal'' domain wall, either bond- 
or site-centered, which is infinitely
extended in $d-2$ directions, and lies along a $45^{\circ}$ angle 
relative to the two remaining lattice directions, which we will call ``x'' 
and ``y''.   As an example, one could construct a
bond-centered diagonal domain wall by placing holes in the
x-y plane along a 
``staircase'' of nearest-neighbor sites obtained by first taking a step in the
x direction, then a step in the y direction, and so on.  To zeroth
order, this diagonal stripe has the same energy per hole as the
vertical, site-centered stripe.

To compare the energies of different sorts of domain walls, we imagine finding
the state of minimum energy of a system of size $2N \times 2N$ in the $x-y$ plane,
and of infinite extent in the remaining $d-2$ directions.  Considering
the projection of the problem on the $x-y$ plane, we study the
possible ground-states of the system with $2N$ holes per plane,
in the presence of
a strong staggered field on the boundary
that favors an up spin on the red sublattice on the 2N sites nearest
the lower corner ({\it i.e.} for points $ x=0, 0 \le y < N $ and
$0 \le x < N, y=0$) and the opposite field,
which favors up-spins on the black sublattice, on the
$4N-2$ sites which form an upper ``cap''
{\it i.e.} the points $x=0, N < y < 2N$, $0 \le x < 2N, y=2N-1$,
and $x=2N-1,N < y < 2N$.  These boundary conditions force the
system to have an antiphase domain wall of length $2N$ sites (or greater),
but permits it to choose whether to have a vertical, diagonal, or
a piecewise vertical domain wall with some concentration of right-angle
kinks.

The kink energy is readily computed perturbatively in powers of $y$ and
$z$, and it also can be re-expressed in powers of $d^{-1}$:
\ba
E_{kink}= &&{J_z \over d}\bigg\{ {y^2(4d^3-13d^2+12d-2)\over 4(d-1)
(2d-3)(4d-5) } \nonumber \\
&&+ {4 z^2(d-3) \over d(2d-3)(d-1)}
+ ... \bigg \}\nonumber \\
=&& {J_z \over 8d}\bigg\{ y^2 + {y^2\over 2d} +{(y^2 +64z^2)\over 4d^2} + {\cal O}(\frac {1} {d^3})    
\bigg \},
\ea
which is manifestly positive in large $d$.  (Here $E_{kink}$
is the energy per site of the $(d-2)$ dimensional hyperline at
which a site-centered vertical domain wall makes a right-angle
bend.) Similarly, the energy of a zig-zag diagonal
domain wall can be compared to the energy of
the vertical wall computed above:
\ba
\epsilon_{diag}-&& \epsilon_{vert} = {J_z(d+1)\over 4d}\bigg[
{dy^2\over(4d-3)(2d-3)(2d-1)(d^2-1)} \nonumber \\
&&+{16z^2\over(d^2-1)(2d-1)} + {\cal O}(z^2y) + {\cal O}(y^4)
+ {\cal O}(z^4)\bigg] \nonumber \\
&&={J_z\over 4d^3}\bigg[ 8z^2 +\frac 1 d \bigg(\frac {y^2} {16}
+12z^2\bigg) + {\cal O}\big(\frac 1 {d^2}\big)\bigg]. 
\ea

Thus it seems that the vertical domain wall has the 
lowest energy in large dimension. However,
it is easy to see that this is a model-dependent result.
For example,
one can readily construct models
on a ``Cu-O'' lattice\cite{cuo}
rather than a hypercubic lattice in which a diagonal, or bond-centered
vertical domain wall has the lower energy.

\subsection{Interactions between two domain walls}

Two domain walls attract each other at long distances through the exchange of 
spin waves, in much the same way as two static holes.  (See above.)
Again, in high dimensions, we expect this effect to be
less important than the short-range attraction
between domain walls.  For instance, to leading order, there is an
attractive interaction energy per unit hyperarea
between nearest-neighbor vertical site centered
domain walls which to leading and next to leading order is simply equal
to the nearest-neighbor attraction between two holes, Eq. (\ref{eq:v2hole}) 
above.

\section{ Behavior in Large but Finite Dimension}

We have found that, in some instances, the electron kinetic energy $t$ 
may play a relatively small role in the physics in large $d$, 
because only states exponentially
near the band minimum have energies of order $-td^0$, while the
bulk of the states have energies of order $t/\sqrt{d}$.   Thus, these
states will only come into play when $t/J$ gets to be parametrically large:
$t/J \sim \sqrt{d}$, where
the large $d$ theory is more difficult to control. In such a regime 
our results are less complete, and more subject to worries that
there could be states we have missed.  For instance, the perturbative
treatment of the one-hole problem is no longer well controlled by large $d$:
As discussed above,
the effective hopping matrix element to the $2n^{th}$ Manhattan
neighbor is of  order $t(t/J)^{2n-1}/d^{3n}$, while the number of such
neighbors increases as $d^{2n}$.  Thus, for $t/J\sim \sqrt{d}$, the contribution
of far neighbor hops to the hole energy for $\vec k$ near 0 or $\vec \pi$
does not decrease with $n$.  Since this only matters for a very small
fraction of one hole-states, while for generic values of $\vec k$, only
the small $n$ terms are important, it is unlikely that this problem leads
to any significant changes in the qualitative physics of the dilute-hole
problem.  It does, however, mean that we cannot be quite as confident
of the completeness of our understanding of the problem in this limit,
as when $t/J$ is not parametrically large.

Nonetheless, with certain plausible assumptions and some guidance from the 
results of various studies in $d=2$, we can elucidate much of the behavior
of the $t-J$ model in this region of parameters, as well.  (Note, as mentioned
previously, our results here are consistent with those obtained
using a somewhat different approach, for the large $d$ Hubbard model.
\cite{dongen})

\subsection{The Phase Diagram for $Jd^{1/2}/t \gg 1$}

To begin with, we consider the behavior of the system when
$J/t\sim 1/\sqrt{d}$ is parametrically small, but $J\sqrt{d}/t>>1$ 
is still large.

For energies away from the band edge, {\it i.e.} for
$\epsilon\sim t/\sqrt{d}$, the density of states per spin
polarization of the
tight-binding model on the hypercubic lattice in large dimension is
readily computed by the method of steepest descents:
\ba
\rho(\epsilon)=&&(2\pi)^{-1}\int dx e^{i x \epsilon} \big[J_0(2tx/d)\big]^d
\nonumber \\
=&& {1\over 2t}\sqrt{d\over \pi} \exp
\big[-\epsilon^2d/4t^2\big]\bigg[ 1 + {\cal O}(\epsilon^2)
\bigg ].
\label{eq:rho}
\ea
From this it is clear that, so long as the electron density is low
enough, we can approximately ignore interactions between electrons, 
({\it i.e.} the
infinite, on-site $U$ 
and the nearest-neighbor antiferromagnetic
interactions), the density as a function of chemical potential 
(for $\mu \le 0$) is readily
seen to be
\be
n(\mu) ={\rm erfc}\big(\sqrt{d\mu^2/4t^2}\big)  + {\cal O}( n(\mu)^2),
\ee
where {\it erfc} is the complementary error function, and the energy density is 
\be
E_{gas}(\mu)= -{2t\over \sqrt{\pi d}} \exp[-d\mu^2/4t^2] + {\cal O}( n(\mu)^2).
\ee

Thus, there is a regime  of parameters, $1 \gg J/t \gg  2/\sqrt{d}$ 
in which the density of {\it electrons} is small (but not
exponentially small) and interactions  between electrons in the ``gas'' phase
can still be neglected in any total energy calculation.  
Under these conditions, $|E_{gas}| \ll |E_{AF}|$ and
$1-n\approx 1$, so from Eq. (\ref{eq:mu}), it is easy to
see that the  density of electrons
in a hole-rich phase in thermodynamic
equilibrium with the undoped antiferromagnet is
\be
n_{max}= {\rm erfc}(J\sqrt{d}/4t) + {\cal O}( n_{max}^2).
\label{eq:nmax2}
\ee

\subsection{The Phase Diagram for $1 \gg Jd^{1/2}/t$}  

When $J/t$ is reduced still further, so that $J\sqrt{d}/t$ gets small,
$n_{max}$ approaches 1, and it is no longer possible to ignore the effects
of interactions in the hole-rich phase. In this limit, we lose the
possiblility of quantitatively reliable results based on our large-$d$
approach.  However, for very small $J/t$ and densities
near 1, it is reasonable to expect the ``electron gas'' state to be 
ferromagnetic, at least locally.  
The ferromagnetic phase is
noninteracting in any dimension when 
$V=-J/4$,  corresponding to the canonical definition of the
$t-J$ model.  In that case, the equilibrium between the ferromagnetic hole-rich
phase and the undoped antiferromagnet can be
computed exactly.  Using the large $d$ expression for the density of states
in Eq. (\ref{eq:rho}), we obtain the  implicit  expression
for $\mu$,
\be
\mu=[E_{AF}-E_{ferro}]/[1-n(\mu)],
\ee
where the ground-state energy of the ferromagnetic Fermi-gas is
\be
E_{ferro} = -{t\over \sqrt{\pi d}} \exp[-d\mu^2/4t^2],
\ee
the electron density is
\be
n(\mu) =(1/2)[2\Theta(\mu)-{\rm sign}(\mu){\rm erfc}\big(\sqrt{d\mu^2/4t^2}\big) ],
\ee
and $\Theta$ is the Heaviside function.  We can evaluate this expression
in the limit of small $J\sqrt{d}/t$:
\be
\mu = {2t\over \sqrt{d}}\sqrt{\ln[2t/\sqrt{\pi d}J}
[1 + {\cal O}(J\sqrt{d}/t)]
\ee
and
\be
1-n_{max}={J\over 2\mu}[1 + {\cal O}(J\sqrt{d}/t)].
\label{eq:nmaxferro}
\ee

This, finally, corrects the unphysical aspect of the phase diagram
for $d\rightarrow \infty$, in that it implies that the boundary of the
two-phase region approaches the zero doping axis (as
$(J\sqrt{d}/t)(\ln[2t/\sqrt{\pi d}J])^{-1/2}$) as $J/t\rightarrow 0$.  
This is shown in the phase diagram in Fig. 2.  (Note that,
because all interaction effects vanish in the ferromagnetic state,
the location of the coexistence line between the ferromagnetic Fermi
liquid and the undoped antiferromagnet can be computed accurately in
any finite dimension $d$, for which $E_{AF}$ is known.\cite{ekl}
For $d=2$, for instance, $1-n_{max}= B\sqrt{J/t}$ where
$B\approx 0.61$.)  

For parametrically small $J/t$ and larger electron
concentration, the nature of the phase diagram in large but finite
dimension is currently unexplored.

\section{The Effect of Coulomb interactions}

The $t-J$ model has been widely studied because it
is supposed to represent the most important low-energy physics of
a system of strongly-interacting charged particles.  
It is assumed that the long-range part
of the Coulomb interaction can be ignored provided it is fairly heavily 
screened by a surrounding dielectric background. But this assumption is
not valid for a state which is macroscopically inhomogeneous.  In the presence
of Coulomb interactions, we need to do thermodynamics 
at fixed mean particle density, and the system must be neutral at
long length scales,  {\it i.e.} phase separation is forbidden.  
In a system with an average electron concentration $1 < n < n_{max}$,
and a long-range but ``weak''
Coulomb interaction in addition to the strong short-range interactions of the
$t-J$ model,  we encounter a class of phenomena that
we have named\cite{frustrated} ``frustrated phase separation''.  Here, the 
system is homogeneous (neutral) on long length scales, but inhomogeneous on 
short length scales, with interleaving regions that look locally like the two 
phases that would coexist in the absence of the Coulomb interactions.  It is 
the purpose of the present
section to explore the consequences of frustrated phase separation in the
$t-J$ model plus ``weak'' long-ranged Coulomb interactions in the limit of 
large $d$.

We define the Coulomb interaction in $d$ dimensions by a generalized 
Poisson equation
\be
-\nabla^2 \phi(\vec r) =\vec \nabla \cdot \vec E = (Q/d) \rho(\vec r),
\ee
where $\phi(\vec r)$ is the scalar potential, $\vec E(\vec r)$
is the electric field, $\rho(\vec r)$ is the particle density, 
\be
U(\vec r)= - (Q/2d) \rho(\vec r)\phi(r) = d\vec E\cdot \vec E/2Q        
\ee
is the energy density, and $Q$ is the effective charge 
(or background dielectric
constant) which determines the strength of the Coulomb interaction.  One could,
of course, imagine different ways of scaling $Q$ in the large $d$ limit.  
Given that
we are after the physics of frustrated phase separation, 
we wish to take the limit
in such a way that i)  macroscopic phase separation is forbidden but 
ii) for
a homogeneous state, the long-range part of the Coulomb interaction is 
unimportant
compared to $t$ and $J$.  This is accomplished by taking the limit in such a 
way that $Q$ does not depend on $d$.

\subsection{An effective model for ``frustrated phase separation''}
In a previous publication, we considered a two dimensional model which
we argued represented the physics of frustrated phase separation at a 
coarse-grained level.  This model is easily generalized to arbitrary dimension:
\ba
H \mbox{    } = -{{\cal J}\over 4}\sum_{<i,j>}(\sigma_i - 1)(\sigma_j - 1) \nonumber
\\  + (1/2)\sum_{i\ne j} V_C(i,j)
[\sigma_i-\bar \sigma] [\sigma_j-\bar \sigma]
\label{eq:frustrated}
\ea
where $\sigma_i=1$ if ``site'' $i$ is a hole-free region and 
$\sigma_j=-1$ if ``site'' $j$ is a hole-rich region, ${\cal J}$ is a short-ranged
``ferromagnetic'' interaction, which promotes macroscopic phase separation
of the two coexisting phases, $V_C$ is the Coulomb interaction, suitably
defined on the lattice, and
\be
\bar \sigma = N^{-1}\sum_j\sigma_j
\ee
is the mean charge per ``site.''  In this model, we imagined that sites represented
small regions which were nonetheless large enough that the local state
could be described as being one of the two phases that would be in
equilibrium with each other in the absence of the Coulomb interaction.  In
the present large dimensional context, it is possible to derive this effective
Hamiltonian microscopically, identifying the sites in the effective model
with the original sites in the $t-J$ model, the $\sigma_i=1$ state with
a site occupied by an electron, the $\sigma_i=-1$ state with an
unoccupied site, and ${\cal J}$ equal to the nearest-neighbor attraction
between two holes, derived in Eq. (\ref{eq:v2hole}) above.  This model is
insensitive to the fact that the hole-rich phase has a non-zero electron
concentration for $J/t < Y_c$, but since the electron concentration is always
exponentially small, this error makes no difference in the energetics and structure
of the various phases of frustrated-separation.  Similarly, it ignores the
fact that in each disconnected region of (hole-free) antiferromagnet, there
is a potential ground-state degeneracy associated with spin-rotational
symmetry;  this might affect the finite temperature behavior of the system,
but has no effect on the ground-state phase diagram.

We have studied\cite{ute} the ground-state phase diagram of this model for
 $d=2$
(and for $V_c(\vec r)= Q/r$, rather than the true 
two-dimensional ``Coulomb'' interaction, $V_c(\vec r)= Q\log[r]$);  the
results there, as well as the results of exact solution\cite{chayes} of a large
N, ``spherical'', version of the same model, lead to the conclusion that
in all dimensions, there are a number of ubiquitous characteristics of the  the
ground-state phase diagram of Eq. (\ref{eq:frustrated}).  Specifically,
for very large $Q/{\cal J}$, the ground state is the ``Wigner crystal'' phase,
or in other words, 
the ground-state of the Coulomb term, itself, which, for $\bar \sigma \le 0$ is a
fully $d$-dimensional crystal of isolated sites with $\sigma=-1$, in a
sea of sites with $\sigma=+1$;  for example, for $\bar \sigma=0$, the
``Wigner crystal'' is a state in which one sublattice is occupied by
$\sigma=+1$ and the other by $\sigma=-1$.  Conversely, for $Q/{\cal J}$
very small, the ground states
consist of a sequence of ``stripe'' phases of varying period (as a function of
$Q/{\cal J}$), or in other words phases in which the density of $\sigma=-1$
is a function of one coordinate, and independent of the remaining
$(d-1)$ coordinates.  In this regime of the phase diagram, for fixed
$\bar \sigma$, 
the smaller $Q/{\cal J}$, the longer the period, as discussed below.
Between the striped phases at small  $Q/{\cal J}$ and the Wigner crystal
phase at large $Q/{\cal J}$, there typically occur a sequence of more complicated
phases that interpolate between the two extremes.  In two dimensions,
we found that these phases occupy an exceedingly narrow sliver of the
phase diagram, but we do not know how generic this behavior is.

\subsection{The properties of the stripe phases}

If we confine ourselves to considerations of stripe phases, then the Hamiltonian
in Eq. (\ref{eq:frustrated}) can be reduced to an effective one
dimensional model by summing over the values of the Ising
spins in the $(d-1)$ dimensions perpendicular to the modulation
direction:
\ba
H_{stripe}=&& -({\cal J}/4d)\sum_j (\sigma_j - 1)(\sigma_{j+1}-1) \\
\label{eq:stripe}
&& +(Q/2d)\sum_{i,j}|i-j|[\sigma_i-\bar \sigma][\sigma_j-\bar\sigma].
\nonumber
\ea

The Madelung energies in this equation involve
infinite sums, which are readily carried out numerically\cite{ute}, but
cannot be done analytically.  However, we can make rather good estimates by replacing
the lattice sums by integrals.  Specifically, for fixed hole concentration,
$x=(1/2)(1-\bar \sigma)$, an array with period $L$,
which consists of alternating  stripes of $\sigma=-1$ of width $W=(L/2)(1-\bar\sigma)$ and
intervening regions of width $(L-W)$ of $\sigma=+1$, is seen in this way to have
energy per unit volume (defined to be volume associated with a single lattice site)
\be
E_{stripe} \approx 
-{{\cal J}W x}+{{QL^2}x^2(1-x)^2\over 24}.
\ee
This expression is readily minimized with respect to $L$, or equivalently the stripe-width
$W$, with the result
\be
W={12{\cal J}\over Q}{x\over (1-x)^2}.
\ee
Finally, recalling that because the lattice is actually discrete, the allowed values of
$W$ are actually nearest integer approximants to this expression, we obtain the approximate
condition that width $n>1$ stripes are stable for
\be
(2n+1) \ge {24{\cal J}\over Q}{x\over (1-x)^2} > (2n-1),
\ee
and that width 1 stripes are stable for 
\be
1 \ge {8{\cal J}\over Q}{x\over (1-x)^2}.
\ee
Of course, for large enough $Q/{\cal J}$, the width 1 stripe phase should give way to
the other phases, mentioned above, and eventually at very large values to the
Wigner Crystal.

An interesting aspect of this model is that, although it only explicitly
involves the enumeration of the charge ordering ({\it i.e.} which sites
are occupied and which are unoccupied), given the fact that there
is an energetic preference for antiphase ordering of the spins across
a charged stripe, we can reconstruct the ground-state spin order (up to
a global rotation) by imposing the constraint that within regions
of occupied sites ($\sigma=1$), the spins are antiferromagnetically
ordered and, across unoccupied sites ($\sigma=-1$), the antiferromagnetic
order suffers a $\pi$ phase shift.  (At very small $Q/{\cal J}$, where the
width of the charged stripe becomes greater than 1, 
a simple generalization of the above microscopic calculations shows
that, in large $d$, the correct ground-state energy and spin order may be 
obtained by viewing a thicker stripe as a collection of nearest-neighbor 
fundamental (width 1) stripes and assigning a $\pi$ phase slip per stripe.)

\section{Generalized Spin Ladders}
There has recently been interest in the properties of spin-1/2 Heisenberg
``ladders'' in $d=2$, where a ladder is effectively a one-dimensional
system which has finite width in all directions, save
one, in which it is infinite.  We can readily apply our analysis to the large 
$d$ generalization
of these ladders.  For instance, we consider a generalized ``two-leg'' ladder,
in which the lattice has a width 2 in $d-1$ directions, and is infinite in
1 direction.   

Proceeding as above, we first obtain a rigorous upper bound,
$E_{Neel}=-(1+1/d)J/8$, and a rigorous lower bound, 
$E_{lower}=-(1+3/d)J/8$ on the ground-state energy per site.
It is interesting, in this context, that the ground-state energy in the
large $d$ limit approaches that of the classical N\'eel state, even
though this is a one dimensional system, so we know rigorously
that there is no true long-range magnetic order.  Indeed, following the
Haldane conjecture,\cite{haldane} it is clear that for any finite $d$,
this system will have exponentially
falling magnetic correlations and a spin gap.  However, this physics
will only be manifest at very long distances (probably exponentially
long in the large $d$ limit), and at shorter distances the system will
appear ordered.

Again, without repeating the earlier analysis, we can compute the ground-state
energy in perturbation theory in powers of $y$, which, to second order,
gives
\be
E_{AF}=-{(d+1) J_z\over 8d} \left[ 1 + {y^2\over d} + 
{y^4(d^2+3d-2)\over 4d^3(d+1)(2d-1)} +{\cal O} (y^6) \right ]
\ee
from which we can deduce that
\ba
E_{AF}= && -{J_z\over 8} \bigg[ 1 + {(1+y^2)\over d} + {y^2\over d^2} + 
{y^4\over 8d^3}  \nonumber \\
&& +{5y^4\over 16d^4} + {\cal O}(1/d^5)
\bigg].
\ea

\section{ Extrapolation to $d=2$ and $d=3$:}
While the $1/d$ expansion gives us a small parameter with which to
analyze the problem, it is legitimate to question  the relevance of 
the $1/d$ results for the physically interesting dimensions $d=2$ and
$d=3$.  (As there exist good methods for solving the present class
of problems in $d=1$, we are not concerned with pushing our results
all the way down to $d=1$.)   

For most of the types of order that we have considered, $d=1$
is the lower critical dimension (for quantum disordering), and it is reasonable
to expect large $d$ results to be qualitatively reliable for 
{\it zero temperature} properties
of the system in $d=2$ and $3$, but not for finite
temperature properties in $d=2$.  
As mentioned previously, this expectation is borne out to
a large extent by comparison of the  
large $d$ phase diagram of the $t-J$ model
shown in Figs. 1 and 2, with the best available data based on
analytic and numerical
studies of the model in $d=2$.

We can get more ambitious, and consider how good the quantitative agreement
is between available numerical and series results for  the model in physical dimensions
and the results of the large dimension expansion.  In Tables 1 and 2 we compare 
known results for the undoped, spin 1/2 Heisenberg model in physical dimensions
with the results of straightforward perturbation theory (in powers of $y=J_{\perp}/J_z$)
and of the $1/d$ expansion.  Clearly, both give quantitatively 
excellent results.  However,
whereas perturbation theory gives its best results if terms only to second order are
retained, the $1/d$ expansion appears to approach closer to the correct value with each
successive order, at least to fourth order (which is the highest order we have computed).
In a sense, for an asymptotic expansion, it is the question of to how high an order
do the results improve, even more than the overall accuracy of the result, which addresses
the issue of how small is the expansion parameter.
\begin{center}
\begin{tabular}{|c|c|c|c|c|c|}
\hline
{} & {$E_{AF}(2)$} & 
{$m(2)$} 
 & {$E_{AF}(3)$} & {$m(3)$} & {$E_{2-leg}(2)$}  \\
\hline
\hline
{$y^0$} & {-0.5} & 
{0.5} & {-0.75} & {0.5} & {-0.375} \\
\hline
{$y^2$} & {-0.6667} 
& {0.3889} & {-0.9} & {0.44} &{-0.5625}\\
\hline
{$y^4$} & {-0.6657} & {0.3929} & {-0.8995} & {0.4404} & {-0.5729} 
 \\
\hline
\hline
{Exact} & {-0.669\cite{2d}} & 
{0.307\cite{2d}} & { - } & { - } & {-0.5780\cite{2leg}}\\
\hline
\hline
\end{tabular}
\end{center}

Table 2 : Comparison of the results of exact numerical studies
(the row labeled ``exact'')
on the two-dimensional spin-1/2 Heisenberg antiferromagnet,
with the perturbative results in powers of $y=J_{\perp}/J_z$ 
derived in the present paper.
The dimension is indicated by the arguments of the computed quantities.  
The approximate results
are obtained by setting $y=1$, $V=0$, and $d=2$ or $3$ in the series expansion,
evaluated to the stated order.  All energies are measured in units of
$J/d$, and the magnetization $m$ is quoted in units in which $g\mu_B=1$,
where $\mu_B$ is the Bohr magneton. 
\vspace*{1cm}

\noindent{\bf Acknowledgments:}
This work was supported in part by the National Science Foundation grant 
number DMR93-12606 (EWC, SAK, and ZN) at UCLA.  Work at
Brookhaven (VJE) was supported by the Division of Materials Science,
U. S. Department of Energy under contract No. DE-AC02-76CH00016.


\begin{references}
\bibitem{tJconvention}  Note that 
the numbers attributed here to Ref. (\cite{onehole})
have been corrected to account for differences in the definition of the 
$t-J$ model.
\bibitem{spingap}  Indirectly, however, the existence of metallic stripe
correlations at intermediate scales, of the sort observed as ordered 
structures in
our large $d$ analysis, is the basis for a recently proposed mechanism
for high temperature superconductivity in the cuprates;  see 
Ref. (\cite{spingap2}).
\end{references}
\end{document}